\documentclass[aps,prb,twocolumn,amsmath,amsfonts,amssymb,showpacs,superscriptaddress,floatfix]{revtex4}
\usepackage{subfigure}
\usepackage{graphicx}
\usepackage{rotating}  
\usepackage{dcolumn}   
\usepackage{times}
\def\bvarepsilon{\boldsymbol{\rule[5.5pt]{0pt}{0pt}\varepsilon}}
\def\bsigma{\boldsymbol{\rule[5.5pt]{0pt}{0pt}\sigma}}
\def\btau{\boldsymbol{\tau}}%
\newlength{\StressFieldWidth}
\newlength{\ScaleWidth}
\newlength{\FieldBoxWidth}
\renewcommand{\atop}[2]{\genfrac{}{}{0pt}{}{#1}{#2}}
\begin{document}
\title{Effective-medium theory for infinite-contrast, 2D-periodic, linear
composites with strongly anisotropic matrix behavior: dilute limit
and cross-over behavior}
\author{Fran\c cois \surname{Willot}}
\email{francois.willot@ensmp.fr} \altaffiliation{Now at:
\emph{Cen\-tre de Mor\-pho\-lo\-gie Ma\-th\'e\-ma\-ti\-que}, 35 rue
St-Honor\'e, F-77305 Fon\-tai\-ne\-bleau Ce\-dex, Fran\-ce.}
\affiliation{D\'epartement de Physique Th\'eorique et Appliqu\'ee,
Commissariat \`a l'\'Energie Atomique, Bruy\`eres-le-Ch\^atel,
F-91297 Arpajon, France.}
\author{Yves-Patrick \surname{Pellegrini}}
\email{yves-patrick.pellegrini@cea.fr} \affiliation{D\'epartement de
Physique Th\'eorique et Appliqu\'ee, Commissariat \`a l'\'Energie
Atomique, Bruy\`eres-le-Ch\^atel, F-91297 Arpajon, France.}
\author{Mart\'\i n I.\ \surname{Idiart}}
\email{mii23@cam.ac.uk} \altaffiliation{Now at: De\-par\-ta\-men\-to
de Aero\-n\'au\-ti\-ca, Fa\-cul\-tad de In\-ge\-ne\-r\'\i a,
U\-ni\-ver\-si\-dad Na\-cio\-nal de La Pla\-ta, Cal\-le 1 y 47,
(1900) La Pla\-ta, Ar\-gen\-ti\-na.}
\author{Pedro \surname{Ponte Casta\~neda}}
\email{ponte@seas.upenn.edu} \affiliation{Laboratoire de
M\'{e}canique des Solides, C.N.R.S. UMR 7649, D\'{e}\-par\-te\-ment
de M\'{e}\-ca\-ni\-que, \'{E}\-co\-le Po\-ly\-tech\-ni\-que, 91128
Pa\-lai\-seau Ce\-dex, France} \affiliation{Department of Mechanical
Engineering and Applied Mechanics, School of Engineering and Applied
Science, University of Pennsylvania, Philadelphia, PA 19104-6315,
USA}
\date{\today}
\begin{abstract}
The overall behavior of a 2D lattice of voids embedded in an
anisotropic matrix is investigated in the limit of vanishing
porosity $f$. An effective-medium model (of the Hashin-Shtrikman
type) which accounts for elastic interactions between neighboring
voids, is compared to Fast Fourier Transform numerical solutions
and, in the limits of infinite anisotropy, to exact results. A
cross-over between regular and singular dilute regimes is found,
driven by a characteristic length which depends on $f$ and on the
anisotropy strength. The singular regime, where the leading dilute
correction to the elastic moduli is an $O(f^{1/2})$, is related to
strain localization and to change in character --- from elliptic to
hyperbolic --- of the governing equations.
\end{abstract}
\maketitle
\section{Introduction}
Effective-medium approximations (EMAs) for nonlinear
composites\cite{ZENG88, BLUM89,
KOTH90,BALL92,PONT92,PONT96,PONT98,PELL01,PONT01,PONT02} (i.e.,
multi-phase materials), which aim to predict their overall (i.e.,
macroscopic) behavior, are pushed to their limits of validity as the
nonlinearity and/or the heterogeneity contrast become
large.\cite{PONT98} Typical examples of this sort of phenomenon in
continuum mechanics include porous,\cite{BISH45,WEIN06,IDIA05} and
rigidly reinforced,\cite{IDIA05, LOPE06} plastic or nonlinearly
elastic media. In the idealized model of perfect plasticity, plastic
material flow takes place at constant stress intensity (the yield
stress). In such circumstances, the flow preferentially concentrates
(\emph{localizes}) in \emph{shear bands}. \cite{KACH04,WRIG02}
Formally, these shear bands are closely related to other types of
minimal breakdown manifolds in heterogeneous media (mechanical
systems as well as nonlinear electrical
networks).\cite{ROUX91,ROUX92,DONE02} However, nonlinear EMAs which
address plasticity rely on a quasi-equilibrium hypothesis, which
means that the characteristic time of an individual ``breakdown''
(or slip) event is longer than that of wave propagation through the
medium\cite{BOKS03} (in nonlinear dielectrics, such conditions are
met as well in the reversible diode network
experiment\cite{BENG88}). Plastic deformation being a strongly
irreversible process, applications of such EMAs to plasticity should
therefore be restricted to \emph{incipient} deformation in an
incremental framework.\cite{WEIN06,KACH04} Nonetheless, upon
neglecting elasticity and assuming plastic incompressibility, the
quasi-equilibrium hypothesis may extend to full-grown deformations,
such as in the slip-line theory of perfect plasticity, but the
governing equations are then hyperbolic.\cite{KACH04}

Efficient nonlinear EMAs rely on the use of an underlying linear
comparison medium,\cite{TALB85,PONT91} which may consist in a
``secant'' (isotropic) approximation\cite{SUQ95} to the nonlinear
response of the composite. In the most recent approaches the
comparison medium is anisotropic, of direction determined by the
applied field,\cite{PONT96} and of strength being consistently
determined by the covariance tensors of the local fields in the
phases.\cite{PELL01,PONT01,PONT02} How these methods cope with
localization at the overall level in heterogeneous media is not well
understood, see Ref.\ \onlinecite{WILL07} and references therein.

To address this issue, this paper is devoted to the signature of
incipient localization in an EMA for periodic
composites.\cite{TORQ05} Because for periodic media efficient
methods of solution have been
developed,\cite{WILL07,NEMA81,NUNA84,TAO85,SANG87,SUQU90,BERG92,MACP94,BURY05}
our focus here is on such materials. Thus, a system consisting of a
two-dimensional (2D) periodic lattice of voids embedded in a
deformable matrix is considered. Aimed at understanding the
hallmarks of localization in the underlying linear medium of
nonlinear EMAs, we focus on the case of an elastically anisotropic
\emph{linear} matrix, of variable anisotropy.\cite{WILL07,OTTO03}

The problem, described in Sec.\ \ref{sec:proform}, admits an exact
analytical solution in the particular case of \emph{infinite}
anisotropy where the governing equations acquire an hyperbolic
character.\cite{WILL07,OTTO03} As a consequence, the overall elastic
moduli depend on the porosity $f$ (the volume fraction of voids) as
powers of $f^{1/2}$, in particular in the dilute limit $f\to
0$.\cite{WILL07} This result is at odds with usual effective-medium
results, in which the first correction to a homogeneous medium is an
$O(f)$,\cite{SANG87,BRUG37} due to its proportionality to the number
of inclusions.\cite{NOTE1}

But for \emph{finite} anisotropy, the governing equations are
elliptic, and no exact solution is available. The crossover to the
regime of high anisotropy and its link to localization, of direct
interest for nonlinear EMAs, and more generally for understanding
the nature of the macroscopic yield transition,\cite{ROUX92} are
investigated hereafter. For lack of exact solutions, comparisons are
made between: (i) quasi-exact numerical results obtained by Fast
Fourier Transform (FFT) calculations; (ii) outcomes of an EMA for
linear periodic media whose predictive capabilities are assessed;
and (iii) the exact results of Ref.\ \onlinecite{WILL07}. The case
of a non-linear (visco-)plastic matrix, of direct experimental
relevance,\cite{WECK06} is examined elsewhere.\cite{IDIA07}

The notation used is as follows: $\mathbb{A}$ denotes a tensor of
components $A_{ijkl}$; the sans-serif $\mathsf{a}$ is the tensor of
components $a_{ij}$ (except for the strain and stress $\bvarepsilon$
and $\bsigma$, and the strain polarization $\btau$, written in
boldface); the boldface $\mathbf{a}$ is the vector of components
$a_i$. A colon denotes a double contraction e.g,
$\mathbb{A}:\mathbb{B}$ has components $A_{ijmn}B_{mnkl}$, etc. For
convenience, indices $i=x$, $y$ or $1$, $2$ are used indifferently
hereafter.

\section{\label{sec:proform}Problem formulation}
\begin{figure}[htbp]
\includegraphics[width=7cm]{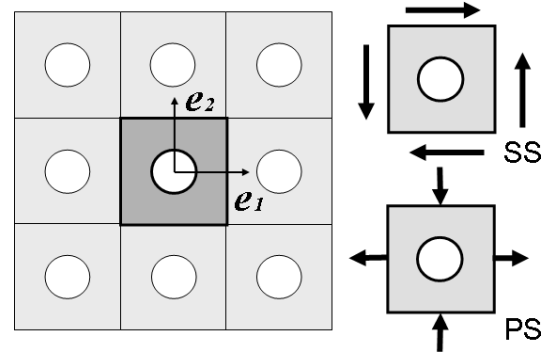}
\caption{\label{fig0} Left, periodic porous medium with unit cell
and reference axes. Right, black arrows depict eigenmodes of strain:
simple shear (SS), and pure shear (PS).}
\end{figure}

The composite, described in Fig.\ \ref{fig0}, consists of an elastic
matrix (phase $\alpha=1$, of volume fraction $c^{(1)}=1-f$),
containing a square array of voided cylinders of radius $a$ (phase
$\alpha=2$, of volume fraction $c^{(2)}\equiv f=\pi a^2$). Here and
henceforth, the size of the unit cell is $L=1$. A set of duality
relations\cite{HELS97} allows one to translate the following results
for the overall behavior of this porous medium, in the context of
rigidly reinforced composites, which is another interesting case of
infinite elastic contrast.

In the composite
$\bsigma(\mathbf{x})=\mathbb{L}(\mathbf{x}):\bvarepsilon(\mathbf{x})$,
where $\mathbb{L}(\mathbf{x})$ is the po\-si\-tion-dependent
elasticity tensor, of components $L_{ijkl}=L_{ijlk}=L_{klij}$. The
equilibrium equation $\partial_i\sigma_{ij}=0$ holds, and the strain
derives from the displacement $\mathbf{u}$ as
$\varepsilon_{ij}=(\partial_i u_j+\partial_j u_i)/2$ (small
perturbations are assumed). In two dimensions,
$\varepsilon_{xx}=\partial_x u_x$, $\varepsilon_{yy}=\partial_y
u_y$, and $\varepsilon_{xy}=(\partial_x u_y+\partial_y u_x)/2$. In
the voids, $\mathbb{L}(\mathbf{x})=\mathbb{L}^{(2)}\equiv 0$, the
stress vanishes, and the strain is arbitrary: any continuation
matching the displacements at the voids boundaries is admissible.
Only its volume average over the void is physically relevant.

The matrix material can be thought of as a ``mixture'' of two basic
types of anisotropic media:\cite{WILL07} (i) one where the
eigendirections of anisotropy coincide with the reference axes of
unit vectors $\mathbf{e}_1\equiv\mathbf{e}^x$ and
$\mathbf{e}_2\equiv\mathbf{e}^y$; and (ii) one where they coincide
with the diagonals (see Fig.~\ref{fig0}). Such a medium is invariant
under the dihedral point-symmetry group $D_4$.\cite{LORM93} Then,
its elastic tensor $\mathbb{L}^{(1)}$ is of the form
\begin{equation}
\label{eq:dqsym} (L_{1111}+L_{1122})\mathbb{J}
+2L_{1212}\mathbb{E}^{\text{SS}}+(L_{1111}-L_{1122})\mathbb{E}^{\text{PS}},
\end{equation}
where $\mathbb{J}$, $\mathbb{E}^{\text{PS,SS}}$ are
mutually orthogonal projectors defined by ($\mathsf{I}$, of
components $\delta_{ij}$, is the $2\times 2$ identity matrix):
\begin{subequations}
\begin{eqnarray}
\mathbb{J}&\equiv&(1/2)\,\mathsf{I}\otimes\mathsf{I},\\
\mathbb{E}^{\text{SS,PS}}&\equiv
&(1/2)\,\mathsf{e}^{\text{SS,PS}}\otimes\mathsf{e}^{\text{SS,PS}}.
\end{eqnarray}
\end{subequations}
The identity is
$\mathbb{I}=\mathbb{J}+\mathbb{E}^{\text{SS}}+\mathbb{E}^{\text{PS}}$.
The definitions involve the so-called \emph{simple shear} (SS) and
\emph{pure shear} (PS) eigenmodes of deformation:
\begin{subequations}
\begin{eqnarray}
\mathsf{e}^{\text{SS}}&\equiv&\mathbf{e}_1\otimes\mathbf{e}_2+\mathbf{e}_2\otimes\mathbf{e}_1,\\
\mathsf{e}^{\text{PS}}&\equiv&\mathbf{e}_1\otimes\mathbf{e}_1-\mathbf{e}_2\otimes\mathbf{e}_2,
\end{eqnarray}
\end{subequations}
such that
$\mathbb{E}^{\text{SS,PS}}:\mathsf{e}^{\text{SS,PS}}=\mathsf{e}^{\text{SS,PS}}$.
Their eigenvectors are related by a $45^{\text{o}}$ rotation (see
Fig.\ \ref{fig0}). Also, $\mathbb{J}:\mathsf{I}=\mathsf{I}$. This
decomposition relates to that of a $2\times 2$ symmetric tensor
$\mathsf{a}$ into one \emph{equibiaxial} (i.e.\ compressive) mode
and two orthogonal shear modes:
\begin{equation}
\label{eq:decomp}
\mathsf{a}=a_m\,\mathsf{I}+a_{\text{SS}}\,\mathsf{e}^{\text{SS}}+
a_{\text{PS}}\,\mathsf{e}^{\text{PS}},
\end{equation}
of respective components $a_{\text{m}}\equiv(a_{xx}+a_{yy})/2$,
$a_{\text{SS}}\equiv a_{xy}$,
$a_{\text{PS}}\equiv(a_{xx}-a_{yy})/2$. Thus, in the matrix we
write:
\begin{equation}
\label{elastmat} \mathbb{L}(\mathbf{x})=\mathbb{L}^{(1)}\equiv 2
\kappa\, \mathbb{J}+2\lambda\, \mathbb{E}^{\text{SS}}+2\mu\,
\mathbb{E}^{\text{PS}}.
\end{equation}
$\kappa$ is the bulk compressibility modulus, and $\lambda$, $\mu$
are in-plane anisotropic shear moduli. With this medium of a special
orthotropic type, the interpretation of the 2D problem as a limiting
one of plane stress ($\sigma_{xz}=\sigma_{yz}=\sigma_{zz}=0$,
$\varepsilon_{zz}\not=0$) or of plane strain
($\varepsilon_{xz}=\varepsilon_{yz}=\varepsilon_{zz}=0$,
$\sigma_{zz}\not=0$) is irrelevant from a theoretical standpoint,
though the expression in terms of $\kappa$ and of $\mu$  of the
Young modulus and Poisson ratio relative to the pure shear mode
differ in both cases.\cite{NOTE2}

With applications to volume-preserving plastic deformation in mind,
this study mostly focuses on the limiting case of \emph{an
incompressible matrix} for which $ \kappa=\infty$. Introducing in
this limit the dimensionless anisotropy ratio $k=\lambda/\mu$, the
medium is isotropic when $k=1$, and is infinitely anisotropic when
either $k=0$ or $k=\infty$. In each of the latter limits, the medium
possesses one infinitely hard, and one infinitely soft eigenmodes of
strain: when $k=0$ (i.e.\ $\lambda=0$ or $\mu=\infty$) the medium is
soft for SS loadings, and resists PS loadings, whereas when
$k=\infty$ (i.e.\ $\lambda=\infty$ or $\mu=0$) the medium is soft
for PS loadings, and resists SS loadings. We accordingly call these
loading modes ``hard'' and ''soft'' hereafter. This model provides a
convenient framework for studying the coupling between porosity and
localization.

Volume averages over the whole system are denoted by brackets
$\langle\cdot\rangle$. Hereafter, $\langle \cdot \rangle^{(\alpha)}$
stands for a volume average over phase $\alpha$. By definition, the
effective (overall) tensor of elastic moduli,
$\widetilde{\mathbb{L}}$, relates the macroscopic strain
$\overline{\bvarepsilon}=\langle\bvarepsilon \rangle$ to the
macroscropic stress $\overline{\bsigma}=\langle\bsigma\rangle$:
\begin{equation}
\label{eq:ltildedef}
\overline{\bsigma}=\widetilde{\mathbb{L}}:\overline{\bvarepsilon}.
\end{equation} The square void lattice also being invariant under
$D_4$, $\widetilde{\mathbb{L}}$ is of a form analogous to
(\ref{elastmat}), where $\kappa$, $\lambda$, $\mu$, are replaced by
the effective moduli $\widetilde{\kappa}$, $\widetilde{\lambda}$,
$\widetilde{\mu}$. The latter are the main quantities of interest.
They depend on $\kappa$, $\lambda$, $\mu$, and $f$. Even when
$\kappa=\infty$, the effective modulus $\widetilde{\kappa}$ is
finite for the porous medium. Then, the normalized moduli
$\widetilde{\lambda}/\lambda$, $\widetilde{\mu}/\mu$ depend only on
$k$, and on $f$. Convenient normalizations for $\widetilde{\kappa}$
are $\widetilde{\kappa}/\mu$ when $\lambda\to\infty$, or
$\widetilde{\kappa}/\lambda$ when $\mu\to\infty$.

\section{\label{sec:fourier}Full-field FFT approach}
\subsection{Numerical method}
Full-field numerical solutions of the problem are obtained using the
Fourier transform method,\cite{MOUL94} applied to linear
comp\-osi\-tes. The method amounts to solving iteratively the
Lippmann-Schwinger equation for the strain,\cite{KORR73}
\begin{subequations}
\label{lippman}
\begin{eqnarray}
\label{eq:int1}
\bvarepsilon(\mathbf{x})&=&\overline{\bvarepsilon}+\int
\text{d}^2\!y\,
\mathbb{G}(\mathbf{x}-\mathbf{y}):\btau(\mathbf{y}),\\
\label{eq:int2} \btau(\mathbf{x})&\equiv&
\left[\mathbb{L}(\mathbf{x})-\mathbb{L}^{(0)}\right]:\bvarepsilon(\mathbf{x}),
\end{eqnarray}
\end{subequations}
where $\mathbb{L}^{(0)}$ is some arbitrary background elastic
tensor. The position-dependent elastic tensor of the medium,
$\mathbb{L}(\mathbf{x})$, is 0 (${}=\mathbb{L}^{(2)}$) in the void
and $\mathbb{L}^{(1)}$ in the matrix. In all the numerical
calculations of the paper, the latter is assumed \emph{nearly}
incompressible with $\kappa\simeq 10^3$, and no appreciable
differences were observed for $\kappa\simeq 10^2$. The tensor
$\mathbb{G}$ is the periodic Green function of the background
medium, such that $\int \text{d}^2\!x\, \mathbb{G}(\mathbf{x})=0$.
In the method, the convolution in Eq.\ (\ref{eq:int1}) is evaluated
in Fourier space, whereas (\ref{eq:int2}) is computed in direct
space. The system is finely discretized as a $L\times L$ array of
pixels. The bad iterative properties of (\ref{lippman}) are
alleviated through various improvements allowing for high or even
infinite contrast.\cite{MICH99,MICH01,MOUL03} These schemes are used
here. Fast convergence is achieved by taking $\mathbb{L}^{(0)}$ of
the type (\ref{elastmat}), with the same anisotropy ratio $k$ as
$\mathbb{L}^{(1)}$, but with considerably lower moduli, namely
$\kappa^0/\kappa\simeq 5.\, 10^{-4}$ and
$\mu^0/\mu\simeq\lambda^0/\lambda\simeq 0.2$ (not necessarily
optimal values). The Fourier transform of $\mathbb{G}$
reads\cite{WILL77}
\begin{equation}
\label{eq:G0} G_{ijkl}(\mathbf{q})=-\left\{q_i
\left[N^{-1}(\mathbf{q})\right]_{jk}q_l\right\}_\text{sym}
\end{equation}
where $\{\cdot\}_\text{sym}$ indicates a symmetrization so that
$G_{ijkl}$ $=$  $G_{klij}$ $=$ $G_{jikl}$, and where
$N_{ij}(\mathbf{q})=q_k\, L^{(0)}_{iklj}\,q_l$ is the acoustic
tensor.

Calculations are carried out for various anisotropy ratios
$10^{-4}\leq k\leq 10^4$, and porosities $0< f< f_c$ using FFT
routines. Three sizes $L=512$, $1024$, $2048$ are considered to
monitor size effects. The smallest one leads to results with
satisfactory convergence properties, except in cases of high
anisotropy where a better resolution was required to render the fine
structure of the field patterns. We used $L=2048$ whenever an
appreciable difference was found between $L=512$ and $1024$.

Once the strain $\varepsilon_{ij}(\mathbf{q})$ is computed, the
displacement
 $\mathbf{u}(\mathbf{q})$ is deduced from ($\mathbf{q}\not=\mathbf{0}$):
\cite{WILL07b}
\begin{eqnarray*}
u_x(\mathbf{q})&=&-i\left\{q_x\left[\varepsilon_{xx}(\mathbf{q})-\varepsilon_{yy}(\mathbf{q})\right]
+2 q_y \varepsilon_{xy}(\mathbf{q})
\right\}/q^2,\\
u_y(\mathbf{q})&=&\hphantom{-}i\left\{q_y\left[\varepsilon_{xx}(\mathbf{q})-\varepsilon_{yy}(\mathbf{q})\right]
-2 q_x \varepsilon_{xy}(\mathbf{q}) \right\}/q^2
\end{eqnarray*}

Only SS or PS \emph{macroscopic} strain loadings are considered
($\overline{\bvarepsilon}_m=0$). Other shear states follow from
linearity. For both modes, the linear elastic problem is solved for
various anisotropy ratios $0\le k\le\infty$. Effective moduli are
computed using one component at a time, e.g.\ $\widetilde{\lambda}
=\langle\sigma_{xy}\rangle/ [2 \langle\varepsilon_{xy}\rangle]$.

\subsection{\label{sec:dispmaps}Overview: displacement and stress maps}
\begin{figure}
\includegraphics[height=8cm,angle=-90]{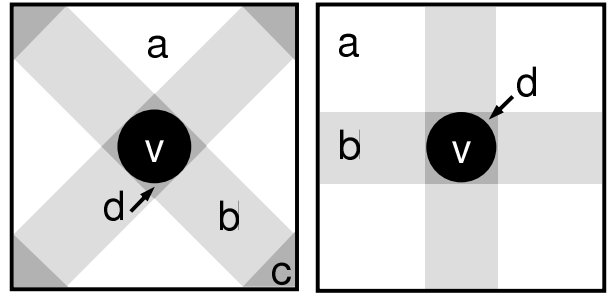}
\caption{\label{fig:bands} \emph{Structure of unit-cell field
patterns for high anisotropy}. Left, pattern for SS loading and
$k=\infty$. Right, pattern for PS loading and $k=0$. In these
figures, \textit{v} $=$ void; \textit{b}, \textit{c}, \textit{d} $=$
deformation bands in the matrix (not intersecting, intersecting far
from the void, and intersecting around the void, respectively);
\textit{a} $=$ remaining parts of the matrix.}
\end{figure}
Typical displacement and stress maps obtained by (isochoric) FFT
calculations are as follows. Since first-order infinitesimal
displacements are used, it should be borne in mind that however
singular, the displacement patterns are at most incipient ones. In
limits of infinite anisotropy $k\to 0,\infty$, the structure of the
solutions tends towards that schematized in Fig.\ \ref{fig:bands},
being organized into \emph{bands} of width one void diameter. Two
remarkable types of structures are found, depending on the loading
direction and on the type of anisotropy. They differ essentially by
the presence of zones in the matrix where the bands cross (denoted
by \textit{c} in Fig.\ \ref{fig:bands}).

With the above mentioned caveat, the following approximate symmetry
holds  between the maps:\cite{WILL07}
\begin{equation}
\label{eq:symetry45} \mathcal{R}_{45^{\text{o}}}(\text{void
lattice}) \Leftrightarrow \left\{
\begin{array}{c}
k\leftrightarrow 1/k\\
\text{PS loading} \leftrightarrow \text{SS loading},
\end{array}
\right.
\end{equation}
where the $\mathcal{R}$ symbol denotes a $45{}^{\text{o}}$ rotation
of the lattice of voids, with all other parameters (material
constitutive law and loading) conserved. These field structures,
already revealed by the analytical calculations of Ref.\
\onlinecite{WILL07} for infinite anisotropy (to which we refer the
reader for further details), are retrieved here for finite, but
high, anisotropy.

\begin{table*}
\caption{\label{tab:displacement} Reduced displacement field
$\mathbf{u}^*$ (arrows), and resulting elastic deformation of the
unit cell (to lowest order of perturbations), in SS and PS loadings
for anisotropy ratios $k=10^{-3}$, $1$, $10^3$. An enlargement of a
void is shown in map (D).}
\includegraphics[width=17.94cm]{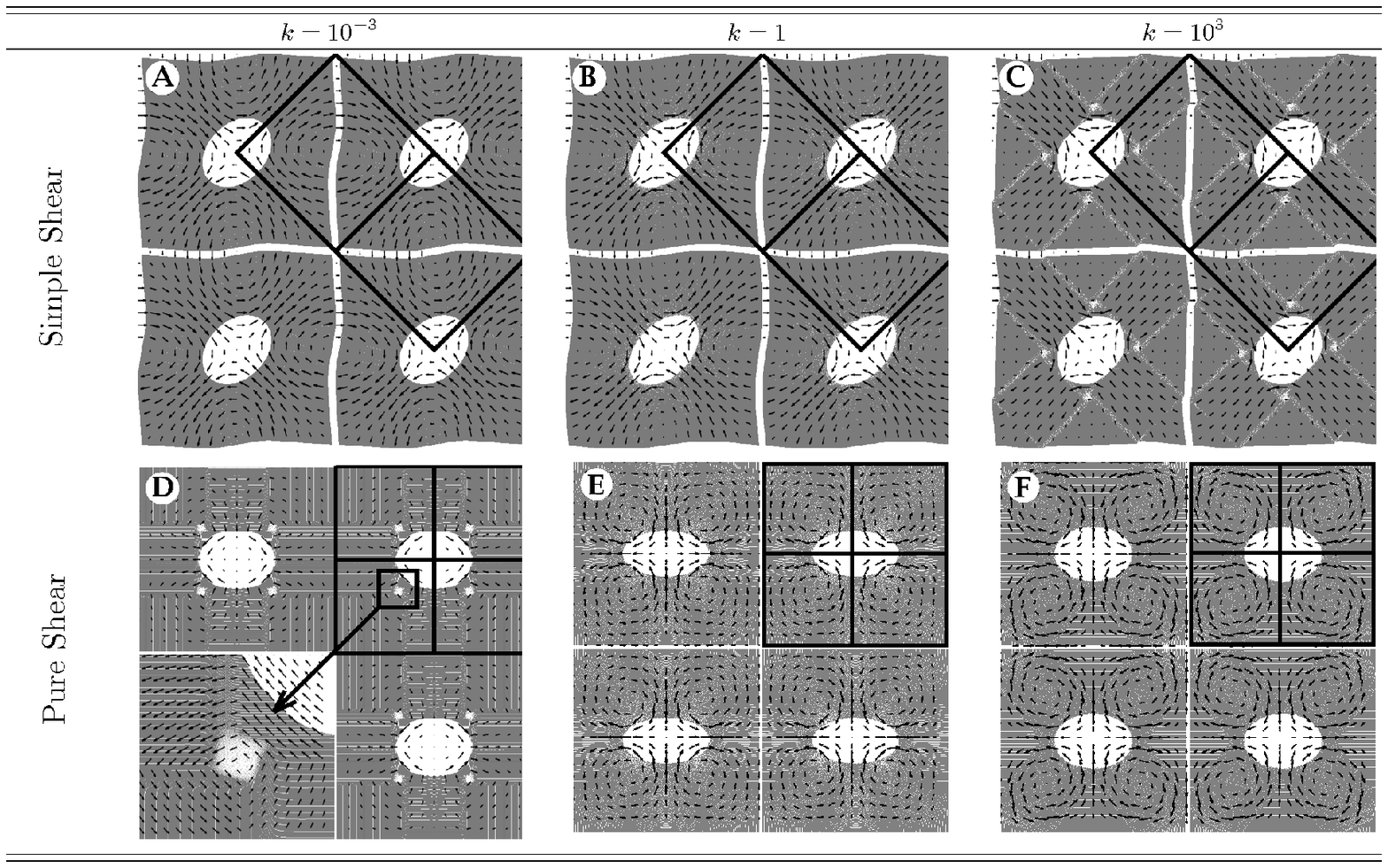}
\end{table*}
Table \ref{tab:displacement} displays full-field calculations of the
reduced (periodic) displacement field
$\mathbf{u}^*(\mathbf{x})\equiv\mathbf{u}(\mathbf{x})-\overline{\bvarepsilon}\cdot\mathbf{x}$,
indicated by arrows, superimposed on a representation of the unit
cell deformed using a rescaled displacement
$\beta\,\mathbf{u}^*(\mathbf{x})$, for anisotropy ratios
$k=10^{-3}$, $1$, $10^3$ in SS and PS loadings, for a moderate
porosity $f=0.1$. To highlight the deformation pattern, the
magnification factor $\beta$ lies between 1 and 10. Lighter grey
tones in the deformed matrix indicate regions subjected to a strong
extension. The unit cell is replicated in order to emphasize the
displacement (``flow'') pattern. For low and high $k$, the features
of the displacement maps are in agreement with the exact results
derived at infinite anisotropy in Ref.~\onlinecite{WILL07}, which
they enlighten.

The flow pattern is organized in closed convection cells of square
shape, delimited by black boxes. Two types of cells, rotated
45${}^o$ with respect to the Cartesian axes, and related by a mirror
symmetry, suffice to account for the flow pattern in SS (maps A, B,
C). As a consequence, and due to the high anisotropy, the edges of
the unit cells in (A) and (C) undergo non-zero and quasi
piecewise-linear deformation. On the other hand, four types of
convection cells, aligned along the Cartesian axes, related by
mirror symmetries with respect to these axes, and fully enclosed
within one unit cell, are required to produce the flow pattern in PS
(D, E, F).

Compared to the $k=1$ isotropic solutions of (B) and (E), solutions
for highly anisotropic situations are either: (i) \emph{localized in
strain}, with a displacement field discontinuous at places [maps (C)
and (D)]; (ii) \emph{localized in stress}, with continuous
displacement as in (A) and (F). Strain-localization arises whenever
loading along a ``hard'' mode takes place. Then, the highly
anisotropic medium resists most the applied strain and undergoes
both a high induced stress and a weak induced strain.  In the limit
of infinite anisotropy, a rigid ``block sliding'' incipient pattern
results, where the flow is organized in bands of width one void
diameter (see also Fig.\ \ref{fig:bands}), where the tangential
component of $\mathbf{u}^*$ is discontinuous, and where strain
concentrates as Dirac distributions along the sliding lines. This
pattern is tantamount to a breakdown mechanism. In turn, block
sliding leaves four incipient voids in the matrix in (C) and (D), at
locations where the sliding lines intersect at 90${}^{\text{o}}$.
One such void is enlarged in (D). One important difference between
cases (C) and (D) is that in (C), the flow bands (of width one void
diameter) cross inside the matrix due to their 45${}^{\text{o}}$
orientation. Flow redistribution then takes place in the
intersection zones. On the contrary, in (D)  such zones do not exist
in the matrix, and flow redistribution requires a non-zero
displacement component normal to the band boundaries. As a result,
the gradient of the tangential component of $\mathbf{u}^*$ is higher
in (C) than in (D).

\begin{table*}[tbp!]
\caption{\label{maps:ss} (Color online) Parallel
($\sigma_\parallel$), transverse ($\sigma_\perp$), and mean
($\sigma_m$) stress field maps for SS and PS loadings, with
anisotropy ratios $k=0.01$, $0.2$, $1$, $5$, $100$ (porosity
$f=0.1$). The stress fields are rescaled such that
$\langle\sigma_\parallel\rangle=1$}
\includegraphics[width=17.94cm]{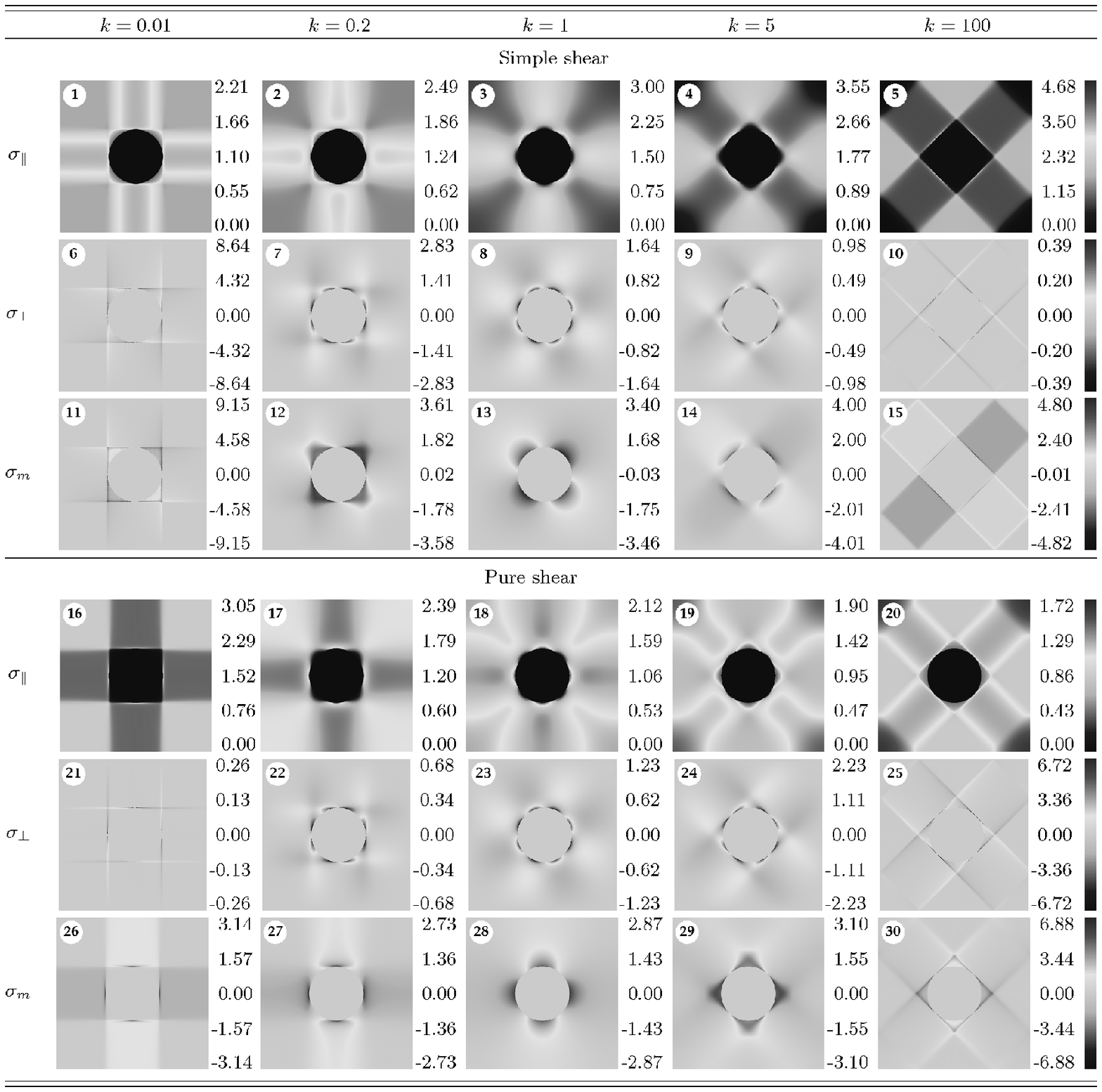}
\end{table*}

Solutions with a continuous displacement field are obtained instead
when loading is applied along the soft deformation mode. The
remaining deformation mode being harder, this leaves less
possibilities for easy deformation than in the isotropic case; this
explains why the unit cells of (A) and (F) are much less deformed --
a magnification $\beta=10$ is used -- than the cells (B) and (E) of
the isotropic material -- plotted with $\beta=1$.

Table \ref{maps:ss} displays, for $f=0.1$ and for increasing
anisotropy ratios $k$, maps of the independent stress components
$\sigma_m$, $\sigma_\parallel$, and $\sigma_\perp$, under SS and PS
loadings. Hereafter, the maps are referred to by their individual
number (1 to 30). Each map goes along with its own field scale at
its right, in correspondence with the color scale at the extreme
right of the rows. The ``parallel'' ($\parallel$) and
``perpendicular'' ($\perp$) notations refer to the ``direction'' of
the applied macroscopic loading. The SS and PS shear components of
the stress are defined in Eq.\ (\ref{eq:decomp}). In PS loading, we
have $\sigma_\parallel\equiv\sigma^{\text{PS}}$,
$\sigma_\perp\equiv\sigma^{\text{SS}}$, whereas in SS loading:
$\sigma_\parallel\equiv\sigma^{\text{SS}}$,
$\sigma_\perp\equiv\sigma^{\text{PS}}$ (herafter, a similar notation
is used for strain components). In both cases, the volume average of
the non-parallel components vanish:
$\langle\sigma_\perp\rangle=\langle\sigma_m\rangle=0$. The maps
display rescaled stresses,
 such that $\langle \sigma_\parallel\rangle=1$. Due to linearity,
the strain fields are the same, up to a change of scale (although
the scales are different in the parallel and perpendicular
directions due to the anisotropy).

The following observations are relevant to the regime of high
anisotropy, where the stress patterns follow that of
Fig.~\ref{fig:bands}. The zones where bands cross depicted in this
figure [either in the matrix (zones \textit{c}) or close to the
voids (zones \textit{d}+\textit{v})], are places of additive
screening or enhancement of the stress. Thus, the parallel stress in
zone \textit{c} of map 20 reaches its highest values there, and is
twice that in the two crossing bands (however, a much higher
transverse stress is encountered in the immediate vicinity of the
void, see map 25). In a similar way, the vanishing stress in zone
\textit{c} in map 5 is the difference between the stresses in the
bands. Two remarks, strictly valid for infinite anisotropy, are in
order at this point: first, zones of vanishing stress are squares,
of size determined by the void cross section transverse to the
bands, so that the disk-like shape of the voids is no longer
relevant; second, the build-up of zones of zero stresses (i.e.,
analogous to porous zones) in the matrix in SS loading leads to an
effective doubling of the porosity in the effective shear modulus
$\widetilde{\lambda}$ at infinite anisotropy, whereby an effective
``close packing'' threshold, twice as small as the geometric
one,\cite{TORQ05} is reached as $f$ increases, leading to a
``mechanically advanced'' percolative behavior. As a consequence,
$\widetilde{\lambda}$ decays rapidly with $f$, see next section.

The stress is less singular than the strain in the limit of infinite
anisotropy. Indeed, in a strain-localized situation (loading along a
``hard'' mode [maps 5, 16]), the displacement is discontinuous.
Accordingly, the transverse strain has Dirac singular components
along the band frontiers. They abruptly change sign at the special
points $(\pm a,0)$ and $(0,\pm a)$ on the void boundary in PS and at
points $(\pm a,\pm a)/\sqrt{2}$ in SS, where $a$ is the void radius.
Because of the stress-strain proportionality, these strain
singularities can be traced in maps 10 and 21. However, since the
perpendicular stress vanishes in the limits $k\to 0,\infty$, so do
its Dirac singularities, as shown by the small values on the scales.
The special points, termed \emph{hot spots} in Ref.\
\onlinecite{WILL07} are points of extreme matter separation, or
crushing, which bear the main cost of the ``block sliding''
patterns. On the other hand, the incipient secondary voids in maps
(C) and (D) of Tab.\ \ref{tab:displacement} appear (somehow
paradoxically) as regions of moderate stress levels.

More generally, the stress field undergoes the following types of
singular
behavior in the limiting cases of infinite anisotropy:\\
--- \emph{loading along a hard mode:}
 discontinuous $\sigma_\parallel$ component along band frontiers
 in the direction normal to the frontiers, with finite jump,
 accompanied by hot spots at the void boundary [maps 5, 16];\\
--- \emph{loading along a soft mode:}
 discontinuous derivative of $\sigma_\parallel$ in the same direction, with
 infinite jump [maps 1, 20], and discontinuous  $\sigma_\perp$ with infinite jump
 across the band frontiers (case of loading along a soft mode) [maps 6, 25].\\
 The mean stress is always singular with the most singular behavior:
it has the singularity of the parallel stress in the case of loading
along a soft mode, and the singularity of the transverse stress in
the case of loading along a hard mode (but the mean strain vanishes
in the limit of an incompressible medium).

\section{\label{sec:ema}Analytical effective medium approach}
Nemat-Nasser proposed\cite{NEMA81} an approximate (dipolar)
Fou\-rier\--mo\-de approach to the periodic problem, which proved
excellent for isotropic components.\cite{NUNA84,SUQU91} We ap\-ply
it to the anisotropic case. More accurate schemes going beyond the
dipolar level, however less suitable to analytical treatment, are
available.\cite{NEMA81,TAO85,BERG92}

The approach is as follows. Consider first the general case of a
binary composite of volume $V\to\infty$, the inclusions of which
have an elastic tensor $\mathbb{L}^{(2)}$, and set
$\delta\mathbb{L}=\mathbb{L}^{(2)}-\mathbb{L}^{(1)}$. The
characteristic function $\chi_\infty$ of an infinite periodic array
of identical inclusions, of characteristic function $\chi$, is
$\chi_\infty(\mathbf{x})=\sum_i \chi(\mathbf{x}-\mathbf{r}_i)$,
where $\mathbf{r}_i$ are lattice vectors. Then,
$\mathbb{L}=\mathbb{L}^{(1)}+\chi_\infty\delta\mathbb{L}$. Equations
(\ref{lippman}), (\ref{eq:G0}) apply, with
$\mathbb{L}^{(0)}=\mathbb{L}^{(1)}$, and
$\btau=\chi_\infty\delta\mathbb{L}:\bvarepsilon$. Multiplying
(\ref{eq:int1}) by $\chi$, integrating over $V$, and assuming
homogeneous deformation in the inclusions so that $\chi_\infty
\bvarepsilon=\chi_\infty\langle
\bvarepsilon\rangle^{(2)}$,\cite{PELL97} results in an expression of
$\langle \bvarepsilon\rangle^{(2)}$ in terms of the Hill
depolarization tensor\cite{HILL65,WILL81} of the lattice
\begin{eqnarray}
\mathbb{P}&=&-\frac{1}{V c^{(2)}}\int{\rm d}^2\!x\,{\rm
d}^2\!x'\,\chi_\infty(\mathbf{x})\mathbb{G}(\mathbf{x}-\mathbf{x}')
\chi_\infty(\mathbf{x}')\nonumber\\
\label{eq:phill1} &=&-V_I \int\frac{{\rm d^2\!
q}}{(2\pi)^2}\sum_{\mathbf{r}}e^{i\mathbf{q}\cdot\mathbf{r}}
\mathbb{G}(\mathbf{q})\left|\left\langle
e^{i\mathbf{x}\cdot\mathbf{q}}\right\rangle_I\right|^2\\
\label{eq:phill}
&=&-c^{(2)}{\sum_{\atop{\scriptstyle\mathbf{q}=2\pi\mathbf{p}}
{\scriptstyle\mathbf{p}\in\text{R.L.}}}\hspace{-1ex}}'
\mathbb{G}(\mathbf{q})\left|\left\langle
e^{i\mathbf{x}\cdot\mathbf{q}}\right\rangle_I\right|^2,
\end{eqnarray}
where $\langle\cdot\rangle_I$ denotes a volume average over
\emph{one} individual inclusion of volume $V_I$. The sum in
(\ref{eq:phill1}) is over lattice sites. The last equality stems
from the Poisson summation formula.\cite{HAUT75} The primed sum is
over nonzero reciprocal lattice vectors $\mathbf{p}=(p_x,p_y)$ (with
integer components). Carrying over the obtained $\langle
\bvarepsilon\rangle^{(2)}$ to the volume average $\langle
\mathbb{L}:\bvarepsilon\rangle$ computed from (\ref{lippman}), and
using definition $(\ref{eq:ltildedef})$, entails the effective
elastic tensor
\begin{equation}
\label{HSLeff} \widetilde{\mathbb{L}} =
\mathbb{L}^{(1)}+c^{(2)}\delta\mathbb{L}:\left(\mathbb{I}+
\mathbb{P}:\delta\mathbb{L}\right)^{-1}.
\end{equation}
The formula for the void lattice with $\mathbb{L}^{(2)}=0$ follows.
%
\begin{figure*}[!htbp]
\includegraphics*[width=17.94cm]{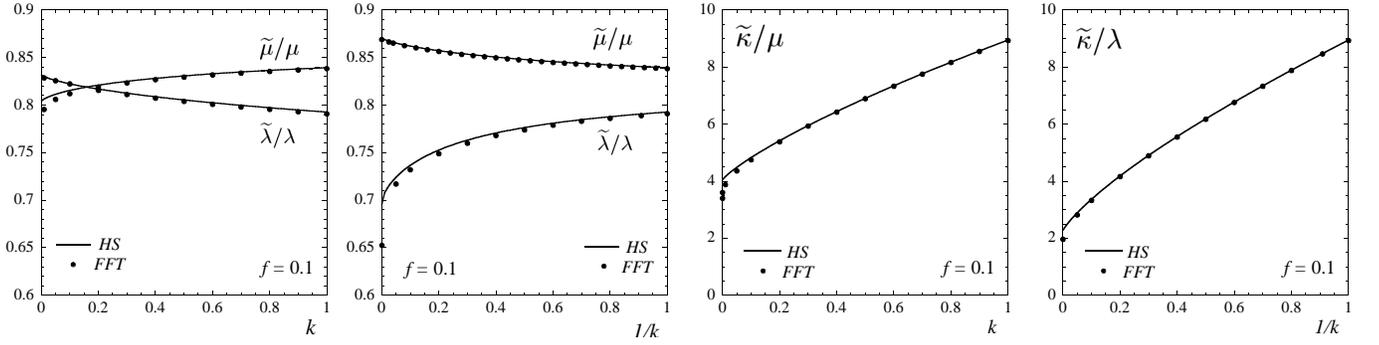}
\caption{Effective shear ($\widetilde{\lambda}$, $\widetilde{\mu}$)
and compressibility ($\widetilde{\kappa}$) moduli vs.\ anisotropy
ratio $k$. Incompressible matrix. Comparisons between the EMA (solid
lines) and FFT results (dots) for porosity $f=0.1$. The quantity
$\widetilde{\kappa}$ is normalized with respect to the most
appropriate modulus, depending on the range of $k$
considered.\label{fig:EffectiveModuliVSk}}
\end{figure*}

Equ.\ (\ref{HSLeff}) is of the Hashin-Shtrikman (HS) variational
type.\cite{HASH63,MILT88,FRIS68} In particular, the void lattice
version provides an upper bound to the exact result.\cite{SUQU91}
Interpreted in the framework of multiple-scattering
theory,\cite{GUBE75,MIDD85} Equ.\ (\ref{HSLeff}) states that, at the
dipolar level, the effect of the inclusion lattice on the
homogeneous matrix amounts to that of \emph{non-interacting}
identical ``equivalent'' inclusions in proportion $c^{(2)}$,
characterized by a $T$-matrix provided by the term following
$c^{(2)}$ in (\ref{HSLeff});\cite{GUBE75,MIDD85} furthermore, each
of these equivalent inclusions possesses a lattice,
$c^{(2)}$-dependent, microstructure, the influence of which is
lumped in $\mathbb{P}$. To make contact with other types of
effective-medium expressions, separate $\mathbb{P}$ into one and
two-body contributions\cite{GUBE75} by writing $\mathbb{P}\equiv
\mathbb{P}_1-c^{(2)}\mathbb{P}_2$, where $\mathbb{P}_1$ is the
$\mathbf{r}=\mathbf{0}$ term in (\ref{eq:phill1}). A similar
decomposition is proposed in Ref.\ \onlinecite{PONT95} in the
context or random composites. Then, introducing
$\delta\widetilde{\mathbb{L}}=\widetilde{\mathbb{L}}-\mathbb{L}^{(1)}$,
(\ref{HSLeff}) takes on the familiar HS form with pair
correlations\cite{WILL77}
$\widetilde{\mathbb{T}}=c^{(2)}\mathbb{T}_1$ where,
\begin{equation}
\label{eq:ttequ}
\widetilde{\mathbb{T}}\equiv\delta\widetilde{\mathbb{L}}:\left(\mathbb{I}+
\mathbb{P}_2:\delta\widetilde{\mathbb{L}}\right)^{-1},\quad
\mathbb{T}_1\equiv\delta\mathbb{L}:\left(\mathbb{I}+
\mathbb{P}_1:\delta\mathbb{L}\right)^{-1}.
\end{equation}
This decomposition proves useful whenever some eigenvalue of
$\mathbb{T}_1$ should blow up. Then, the corresponding eigenvalue of
$\delta\widetilde{\mathbb{L}}$ is simply provided by that of
$-\mathbb{P}_2^{-1}$, as the above expression makes clear.

For cylindrical voids of radius $a$, with $J_1$ the Bessel function,
$\langle e^{i\mathbf{q}\cdot\mathbf{x}}\rangle_I$ $=$$ 2J_1(a q)$
$/(a q)$. Setting
\begin{equation}
\label{eq:mldef}
m\equiv\mu/\kappa,\quad \ell\equiv\lambda/\kappa,
\end{equation}
one finds from (\ref{eq:G0}), (\ref{eq:phill}) that
$P_{ijkl}=\{Q_{ipql}R_{jpqk}\}_{\rm sym}$, where $\vphantom{\bigl(}
\mathbb{R}=\mathbb{J}-(1+m)\mathbb{E}^{\text{SS}}-(1+\ell)\mathbb{E}^{\text{PS}}$,
where ``sym'' denotes a symmetrization with respect to indices
$(i,j)$ and $(k,l)$, and where
\begin{subequations}
\begin{eqnarray}
\mathbb{Q}=\frac{2}{\pi}{\sum_{\mathbf{p}\in\text{R.L.}}\hspace{-1ex}}'\frac{J_1^2(2\pi
a p
)}{p^2\Delta(\mathbf{p})}\mathbf{p} \otimes\mathbf{p}\otimes\mathbf{p}\otimes\mathbf{p},\\
\Delta(\mathbf{p})=\lambda(1+m)\left(p_x^2-p_y^2\right)^2+4\mu(1+\ell)p_x^2
p_y^2.
\end{eqnarray}
\end{subequations}
The reciprocal lattice is a square lattice. Hence $\mathbb{Q}$ is
also invariant under $D_4$. Being completely symmetric, it is of
type (\ref{eq:dqsym}) with $L_{1122}=L_{1212}$ and is determined by
two independent scalar lattice sums only. One obtains:
\begin{equation}
\mathbb{Q}=\frac{1}{\mu
(1+\ell)}\left[(S_\lambda+S_\mu)\mathbb{J}+S_\mu\mathbb{E}^{\text{SS}}+S_\lambda\mathbb{E}^{\text{PS}}\right],
\end{equation}
where, after having reduced the lattice sums to sums over the
positive quadrant,
\begin{equation}
\label{eq:ssums} \left. \genfrac{}{}{0pt}{0}{S_\lambda}{S_\mu}
\right\} =\frac{4}{\pi}\sum_{\genfrac{}{}{0pt}{1}{p_x\geq 0}{p_y\geq
1}} \frac{J_1^{\,2}(2\pi a p)} {p^2\left[4 p_x^2 p_y^2+k
\left(p_x^2-p_y^2\right)^2\right]} \left\{
\genfrac{}{}{0pt}{0}{\left(p_x^2-p_y^2\right)^2}{4 p_x^2 p_y^2}
\right. .
\end{equation}
These sums bring in the anisotropy parameter:
\begin{equation}
\label{eqkdef} k\equiv [(1+m)\lambda]/[(1+\ell)\mu]
\end{equation}
which reduces to $\lambda/\mu$ in the incompressible limit
$\kappa\to\infty$. We remark in passing that
\begin{equation}
\label{eq:relation} k
S_\lambda+S_\mu=S_2(a)\equiv\frac{4}{\pi}\sum_{p_x\geq 0, p_y\geq 1}
\left[J_1(2\pi a p)/p\right]^2
\end{equation}
is independent of $k$. After some algebra, one arrives at
\begin{eqnarray}
\mathbb{P}\!\!&=&\!\!\frac{1}{2\mu (1+\ell)}\left\{(\ell S_\lambda+
m S_\mu)\mathbb{J}+[m S_\mu+(1+m)S_\lambda]\mathbb{E}^{\text{SS}}\right.\nonumber\\
\label{eq:ptens} &&{}+\left.[\ell
S_\lambda+(1+\ell)S_\mu]\mathbb{E}^{\text{PS}}\right\}.
\end{eqnarray}

The one-body $\mathbb{P}_1$ is read from this expression, provided
that $S_{\lambda,\mu}$ are computed in the continuum limit, by
making the substitutions
$\sum\to\frac{1}{4}\lim_{\epsilon->0}\int_\epsilon^\infty {\rm
d}^2\!q/(2\pi)^2$, $\mathbf{p}\to\mathbf{q}/(2\pi)$ in
(\ref{eq:ssums}). Then (in the continuum limit),
$S_{\lambda,\mu}\to$
\begin{equation}
\label{eq:sonebody} S_{1\mu}\equiv\frac{1}{1+\sqrt{k}},\qquad
S_{1\lambda}\equiv\frac{1}{(1+\sqrt{k})\sqrt{k}}.
\end{equation}
Eqs.\ (\ref{eq:ssums}) show that $S_\lambda$ blows up when $k\to 0$
due to the contribution of the Cartesian axis $p_x=0$. On the other
hand, $S_\mu$ remains finite or goes to zero in all cases.

The limit of an isotropic matrix where $k=1$, $\ell=m=\kappa/\mu$
provides $\mathbb{P}_1=[2m\mathbb{J}+(1+2m)\mathbb{K}]/[4\mu(1+m)]$,
where
$\mathbb{K}\equiv\mathbb{E}^{\text{SS}}+\mathbb{E}^{\text{PS}}$.
This expression can be recovered directly from (\ref{eq:G0}) and
from the usual definition in terms of an angular
integral\cite{KNEE65} $\mathbb{P}_1=-\int {\rm
d}^2\!\Omega_{\mathbf{\hat q}}\mathbb{G}(\mathbf{\hat q})/(2\pi)$
where $\mathbf{\hat q}=\mathbf{q}/q$ (the independence wrt.\ $\chi$
stems from the rotational symmetry of the voids).

From (\ref{HSLeff}), (\ref{eq:ptens}), the effective moduli of the
void lattice read, with $f=c^{(2)}$:
\begin{subequations}
\label{eq:effmods}
\begin{eqnarray}
\label{eq:kappaeff}
\hspace{-1em}\widetilde{\kappa}/\kappa&=&
1\!-\!f/\left\{1-[(\lambda/\mu)S_\lambda+S_\mu]/(1+\ell)\right\},\quad\\
\label{eq:lambdaeff} \hspace{-1em}\widetilde{\lambda}/\lambda&=&
1-f/\left\{1-k[S_\lambda+m S_\mu/(1+m)]\right\},\quad\\
\label{eq:mueff} \hspace{-1em}\widetilde{\mu}/\mu&=&
1-f/\left\{1-[S_\mu+m S_\lambda/(1+\ell)]\right\}.\quad
\end{eqnarray}
\end{subequations}
Henceforth, incompressibility is assumed for simplicity so that
$k\equiv\lambda/\mu$ from now on, unless explicitly stated.
\section{Results}
\subsection{\label{sec:effmod}Effective moduli}
The numerical results at various values of $k$ and $f$ discussed in
this section are obtained using brute force numerical computations
of the sums $S_{\lambda,\mu}$, with convergence checks. The sums
$S_{\lambda,\mu}$ are conditionally (and slowly) convergent and the
following suitable prescription is used. Sums are carried out over
concentric square shells of points $\mathcal{S}_n=\{(p_x,n)|\,0\leq
p_x\leq n-1\}\cup\{(n,p_y) |\,1\leq p_y\leq n\}$ for $1\leq n\leq
N$, with $N$ is sufficiently large. Huge numbers of terms are
required for accuracy, especially in the dilute limit.

Figs.~\ref{fig:EffectiveModuliVSk} show comparisons between the
effective moduli computed numerically from the above maps, and the
EMA of Sec.\ \ref{sec:ema} (indicated as HS in the plots), for
$f=0.1$. The agreement is excellent near the case of an isotropic
matrix $k=1$ (as is expected for such a small porosity), but also up
to high anisotropy. In all cases, the EMA is seen to provide an
upper bound for the corresponding full-field estimates (a property
of the Hashin-Shtrikman approach).
%
\begin{figure}[!tbp]
\includegraphics*[width=7cm]{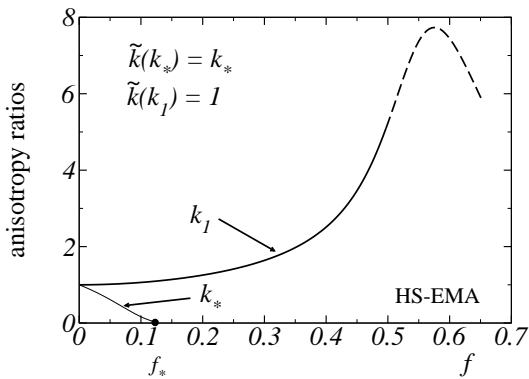}
\caption{\label{fig:SpecialAnisotropyRatios}Anisotropy ratios $k_1$
and $k_*$ vs. porosity $f$ in the HS-EMA. Incompressible matrix.}
\end{figure}
%
\begin{figure*}[!tbp]
\includegraphics*[width=17.94cm]{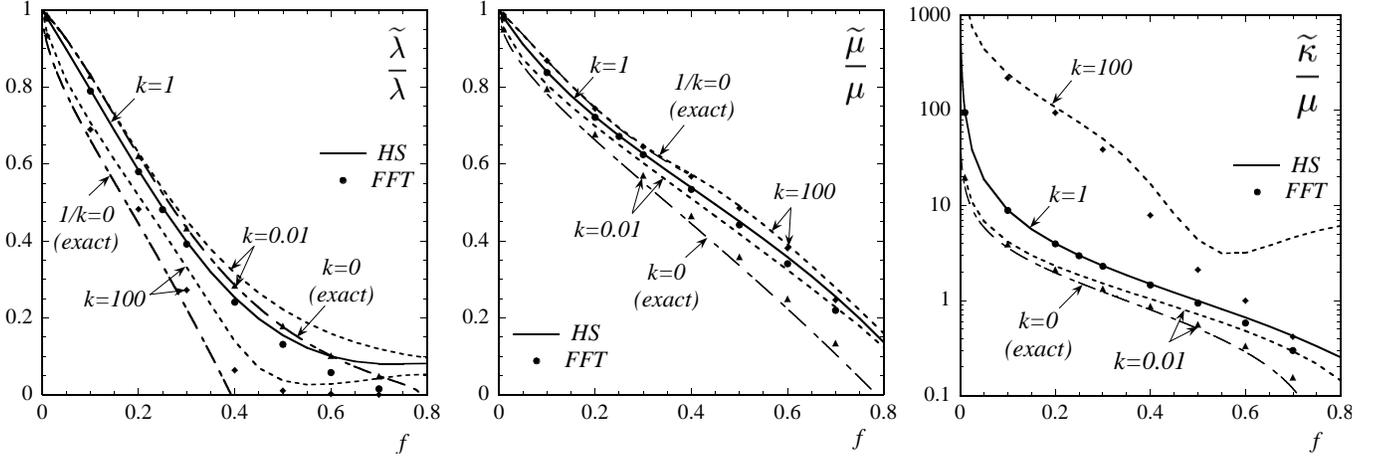}
\caption{\label{fig:EffectiveModuliVSc} Effective shear
($\widetilde{\lambda}$, $\widetilde{\mu}$) and compressibility
($\widetilde{\kappa}$) moduli vs.\ porosity $f$ for various
anisotropy ratio $k$.  Incompressible matrix. Comparisons between
the EMA for $k=1$ (solid), and $k=0.01$, $100$ (dotted); FFT results
for $k=0.01$ (triangle dots), $k=0.01$ (circle dots), and $k=100$
(losange dots); and exact results at $k=0$, $\infty$ taken from
Ref.\ \onlinecite{WILL07} (dash-dotted).}
\end{figure*}

Near $k=1$, the orientation of the void lattice makes the medium
harder under PS loading, than under SS loading [i.e.,
$\widetilde{\mu}(k=1)$ $>$ $\widetilde{\lambda}(k=1)$]. Indeed, the
anisotropic matrix can be thought of as containing rigidifying
fibers (of strength $\mu$), oriented at 45${}^o$ along the
diagonals, that resist PS deformation, and fibers (of strength
$\lambda$) oriented along the Cartesian axes, that resist SS
deformation. In the void lattice, the nearest neighboring voids, and
consequently the largest ``directional damage'', lie along the
Cartesian axes, which explains the difference. We emphasize that
while this observation remains true as $k\to 0$, the situation
changes as $k\to\infty$: in this limit, due to the presence of the
fictitious voids produced by band crossing, mentioned in the
previous section, the nearest-neighboring ``voids'' become located
along the diagonals, so that the PS direction becomes, for $k$
higher than some value $k=k_1$ (discussed below), the most damaged
one, hence the softest.

As $k\to 0$ (Fig.~\ref{fig:EffectiveModuliVSk}a), the curvature of
the plots indicates that $\widetilde{\mu}$ increases slower than
$\mu$, whereas $\widetilde{\lambda}$ decreases slower than
$\lambda$. As $k\to\infty$ (Fig.~\ref{fig:EffectiveModuliVSk}b), the
parts played by $\lambda$ and $\mu$ are reversed. Moreover, the
compressibility modulus $\widetilde{\kappa}$  decreases as
anisotropy increases, in a way comparable to the hardest shear
modulus (Figs.~\ref{fig:EffectiveModuliVSk}c and d).

At high anisotropy $k,1/k \lesssim 0.1$, discrepancies between
full-field calculations and the EMA arise for the hardest shear
modulus (i.e., $\widetilde{\mu}$ when $k\to 0$, and
$\widetilde{\lambda}$ when $k\to\infty$), whereas the softest one
remains extremely well reproduced. This may indicate that the
lattice sums have problems dealing accurately with the effect of
second-nearest-neighboring voids. Indeed, the softest direction is
always the one where the voids (real, or fictitious) are
nearest-neighbors, whereas the hardest one corresponds to
second-nearest-neighbors. Note that $\widetilde{\kappa}$, which
behaves as the hardest effective modulus, suffers similar
discrepancies at high anisotropy.

To discuss the crossing of the curves that takes place in
Fig.~\ref{fig:EffectiveModuliVSk}a, consider the effective
anisotropy ratio
$\widetilde{k}\equiv\widetilde{\lambda}/\widetilde{\mu}$. Crossing
occurs when $\widetilde{k}(k_*)=k_*$, for some $k=k_*(f)$ where the
overall medium and the matrix have the same anisotropy ratio. The
point $k_*$, as estimated by the EMA, is represented vs.\ $f$ in
Fig.~\ref{fig:SpecialAnisotropyRatios} (no attempt has been made to
use full-field calculations for computational cost reasons). The
$k_*(f)$ curve shows that crossing only occurs for porosities
$f<f_*\simeq 0.13$: as $f$ increases, the curve
$\widetilde{\lambda}/\lambda$ in
Fig.~\ref{fig:EffectiveModuliVSk}(a) goes down to zero faster than
$\widetilde{\mu}/\mu$, while the crossing point shifts to the left
until it vanishes. Remark that $k_*<1$ whenever it exists. For
$k<k_*<1$, the matrix is more anisotropic than the composite; the
inverse situation prevails for $k>k_*$, and in particular for
$f>f_*$ where $k_*\equiv 0$, so that void-induced anisotropy
dominates in this regime.

The other remarkable anisotropy ratio is the aforementioned $k_1$,
defined by the equation $\widetilde{k}(k_1)=1$, where the overall
behavior is isotropic in the plane. This point, also represented on
Fig.~\ref{fig:SpecialAnisotropyRatios}, exists at least up to high
porosity values. However, since the EMA is expected to fail around
$f=0.5$ (see below), the irrelevant part of the $k_1(f)$ curve is
sketched with dashed lines in
Fig.~\ref{fig:SpecialAnisotropyRatios}. In the relevant porosity
range, the fact that $k_1(f)>1$ indicates that the matrix needs to
be made harder along the SS ($\lambda$) mode than along the PS
($\mu$) mode in order to reach isotropy, so as to compensate for
higher softening in this direction due to newly appearing nearest
neighboring voids, as is explained above.

Fig.\ \ref{fig:EffectiveModuliVSc} illustrates the behavior of the
moduli with the porosity $f$, for finite anisotropy ratios $k$ $=$
$0.01$, $1$, $100$, together with the exact results of Ref.\
\onlinecite{WILL07} at $k=0$, $\infty$. The exact curve for
$\widetilde{\mu}$ at $k=\infty$, almost superimposed with the EMA
curve for $k=100$ in (b), is available up to $f=\pi/8$
only.\cite{WILL07} Firstly, the EMA is again seen to systematically
overestimate the moduli. Next, all the elastic moduli must vanish at
least at the geometrical close-packing threshold of the
voids,\cite{TORQ05} $f=f_c=\pi/4\simeq 0.78$, and possibly
before.\cite{WILL07} The FFT points in
Figs.~\ref{fig:EffectiveModuliVSc}a and
\ref{fig:EffectiveModuliVSc}b are consistent with this fact, whereas
the EMA fails by producing non-zero results at this point. This is
not surprising, since EMAs of the HS type are known not to be able
to account for percolative-type behavior.\cite{TORQ05} Moreover, the
exact result in Fig.\ \ref{fig:EffectiveModuliVSc}a for $k=\infty$
shows the shear modulus in the hard direction,
$\widetilde{\lambda}$, to vanish at $f=f_c/2$ due to the fictitious
voids produced by band crossing. Accordingly, for large but finite
$k$, $\widetilde{\lambda}$ decreases rapidly with $f$ up to
$f=f_c/2$, then with a lower slope up to $f=f_c$. The EMA again
fails to account for the threshold at $f_c/2$, although the local
minimum of $\widetilde{\lambda}$ at $f\simeq 0.55$ in Fig.\
\ref{fig:EffectiveModuliVSc}a may indicate that at least part of the
phenomenon is captured by the dipolar lattice sums. Interestingly
enough, when available, the exact results for infinite anisotropy at
$k=0$ (resp.\ $k=\infty$) are seen to provide tight lower (resp.\
upper) bounds to the effective moduli for all values of $k$, and in
particular to the isotropic case $k=1$. As far as the effective
moduli are concerned, Fig.\ \ref{fig:EffectiveModuliVSc} clearly
shows that the EMA can be trusted quantitatively up to $f=0.30$ at
most, and is qualitatively reasonable (as long as the matrix is not
too anisotropic) up to $f=0.5$.

\subsection{\label{sec:dilim} Continuous transition in the dilute limit $f\ll 1$}
\subsubsection{Finite anisotropy}
\label{sec:fian} For a finite anisotropy ratio $k$, the dilute
expressions for the effective \emph{shear} moduli at sufficiently
small $f$ are read from expressions (\ref{eq:lambdaeff}),
(\ref{eq:mueff}) with $S_{\lambda,\mu}$ replaced by the one-body
contributions $S_{1\lambda}$ and $S_{1\mu}$ defined in
(\ref{eq:sonebody}). For the incompressible medium, the HS estimates
of the shear moduli are:
\begin{subequations}
\label{eq:mu0PS}
\begin{eqnarray}
\label{eq:lambkfin}
\widetilde{\lambda}/\lambda&=&1-f(1+\sqrt{k})+O\left(f^2\right),\\
\label{eq:mukfin}
\widetilde{\mu}/\mu&=&1-f(1+1/\sqrt{k})+O\left(f^2\right).
\end{eqnarray}
As to the effective compressibility modulus, the incompressible
limit leads to the situation described below Equ.\ (\ref{eq:ttequ}),
where one eigenvalue of $\mathbb{T}_1$ blows up. This requires us to
go beyond the one-body approximation. However, Equ.\
(\ref{eq:asympt2}) in the Appendix \ref{sec:latticesums} shows that
$S_2(a)$ in (\ref{eq:relation}) is \emph{exactly} $S_2=1-f$ for
$f<\pi/4$. Replacing, e.g.\ $S_\mu$ by $S_{1\mu}+O(f)$ and computing
$S_\lambda$ via (\ref{eq:relation}), then letting $\kappa\to\infty$
in (\ref{eq:kappaeff}) [with $k$ read from (\ref{eqkdef})] provides:
\begin{equation}
\label{eq:kappakfin} \widetilde{\kappa}=\sqrt{\lambda\mu}/f+O(1).
\end{equation}
\end{subequations}
Remark that ``extended" dilute approximations, which extrapolate the
above formulas for moderate anisotropy to finite (but small)
porosities, result from taking $\mathbb{P}_2=\mathbb{P}_1$ in
(\ref{eq:ttequ}), i.e.\ from using in (\ref{eq:effmods}):
\begin{equation}
\label{eq:extdil}
S_{\mu,\lambda}=S_{\mu,\lambda}^{\rm
dil}\equiv(1-f)S_{1\,\mu,\lambda}.
\end{equation}
This amounts to assuming pair correlations between the voids
dictated by the void shape,\cite{PONT95} and provides HS formulas of
the ``classical" type in which the lattice structure is ignored.
%
\begin{figure}[ht!]
\includegraphics[width=8.5cm]{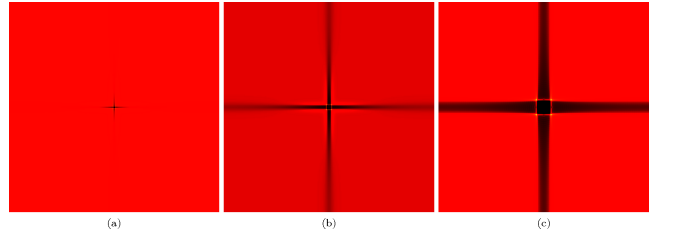}
\caption{ \label{fig:chgmtregFFT} (Color online) FFT computations of
$\varepsilon_\text{PS}=\varepsilon_\parallel$ in PS loading, for a
material with strong anisotropy ratio $k=10^{-3}\ll 1$. Porosities
from left to right:
 $f=f_r/10$ (regular ``dilute'' regime);
 $f=f_r$ (crossover regime); $f=10 f_r$ (``strongly anisotropic'' regime), where $f_r\simeq
k/\pi$ is the cross-over porosity. Incompressible matrix. Black
represents the highest field values (arbitrary color scale).}
\end{figure}
%
\subsubsection{Infinite anisotropy}
After the incompressible limit $\kappa\to\infty$ is taken, the limit
of infinite anisotropy $k\to 0$ is obtained by letting $\mu\to
\infty$ in $\widetilde{\lambda}$, and $\lambda\to 0$ in
$\widetilde{\mu}$. Conversely, $k\to\infty$ requires $\mu\to 0$ in
$\widetilde{\lambda}$, and $\lambda\to \infty$ in $\widetilde{\mu}$.
In these limits, the sums $S_\lambda$, $S_\mu$ in (\ref{eq:ssums})
are computed in the Appendix \ref{sec:latticesums}. The obtained
shear moduli $\widetilde{\lambda}$, $\widetilde{\mu}$ are compared
to the exact results of Ref.\ \onlinecite{WILL07}. One finds for
$k\to 0$:
\begin{subequations}
\label{eq:kzerlim}
\begin{eqnarray}
\label{eq:lambzer}
\frac{\widetilde{\lambda}}{\lambda} &=&1-f-\frac{32}{3}\left(\frac{f}{\pi}\right)^{3/2}\!\!+2\left(1-\frac{512}{9\pi^3}\right)f^2+O(f^{5/2}),\nonumber\\
\\
\label{eq:muzer}\frac{\widetilde{\mu}}{\mu} &=&
1-\frac{3\pi^2}{32}\left(\frac{f}{\pi}\right)^{1/2}-\frac{9\pi^3}{1024}f+O(f^{3/2}),\\
\label{eq:kazpzer}\frac{\widetilde{\kappa}}{\mu} &=&
\frac{32}{3\pi^2}\left(\frac{\pi}{f}\right)^{1/2}-2.
\end{eqnarray}
while exact expressions are:
\begin{eqnarray}
\frac{\widetilde{\lambda}}{\lambda} &=&
1-f-\frac{32}{3}\left(\frac{f}{\pi}\right)^{3/2}
\hspace{-1em}+\left(1-\frac{6}{\pi}-\frac{8}{\pi^2}\right)f^2+O(f^{5/2}),\nonumber\\
\label{eq:lambzerex}
\\
\label{eq:muzerex}\frac{\widetilde{\mu}}{\mu} &=&
1-\left(\frac{f}{\pi}\right)^{1/2}.
\end{eqnarray}
\end{subequations}
For $k\to \infty$, the HS estimates read:
\begin{subequations}
\label{eq:kinflim}
\begin{eqnarray}
\label{eq:lambinf} \frac{\widetilde{\lambda}}{\lambda} &=&
1-\frac{3\pi^2}{16\sqrt{2}}\left(\frac{f}{\pi}\right)^{1/2}\hspace{-1em}-\frac{9\pi^3}{512}\left(\frac{f}{\pi}\right)+O(f^{3/2}),\\
\frac{\widetilde{\mu}}{\mu} &=&
1-f-\frac{16\sqrt{2}}{3}\left(\frac{f}{\pi}\right)^{3/2}\hspace{-1em}+2\left(1-\frac{256}{9\pi^3}\right)\left(\frac{f}{\pi}\right)^2\nonumber\\
\label{eq:muinf}
&&{}\hspace{4.5cm}+O(f^{5/2}),\\
\label{eq:kapinf}\frac{\widetilde{\kappa}}{\lambda} &=&
\frac{16\sqrt{2}}{3\pi^2}\left(\frac{\pi}{f}\right)^{1/2}-2.
\end{eqnarray}
whereas exact expressions are:
\begin{eqnarray}
\label{eq:lambinfex}
\frac{\widetilde{\lambda}}{\lambda} &=& 1-\left(\frac{2f}{\pi}\right)^{1/2},\\
\label{eq:muinfex}\frac{\widetilde{\mu}}{\mu} &=&
1-f-\frac{32}{3\sqrt{2}}\left(\frac{f}{\pi}\right)^{3/2}
\hspace{-1em}+\left(1-\frac{3}{\pi}-\frac{4}{\pi^2}\right)f^2.
\end{eqnarray}
\end{subequations}
The above comparisons show that the HS estimates do an excellent job
of capturing the presence of half-integers powers of $f$ in limits
of infinite anisotropy at lowest orders in the dilute limit.
Moreover, even when the numerical coefficients are not exact, they
are close to the exact values. The less singular character of
$\widetilde{\lambda}$ in (\ref{eq:lambzer}), (\ref{eq:lambzerex})
when $k\to 0$ [resp.\ $\widetilde{\mu}$ in (\ref{eq:muinf}),
(\ref{eq:muinfex}) when $k\to\infty$] is discussed in Ref.\
\onlinecite{WILL07}.

\subsubsection{The dilute transition}
\label{sec:tdt} Obviously, a cross-over takes place between sets
(\ref{eq:mu0PS}) on the one hand, and (\ref{eq:kzerlim}),
(\ref{eq:kinflim}) on the other hand. Balancing the ``extended
dilute" sum $S^{\rm dil}_\mu$ (\ref{eq:extdil}) with $S_\mu^{k\to
0}$ [equ.\ (\ref{eq:limitsums2})], then with $S_\mu^{k\to \infty}$
[equ.\ (\ref{eq:limitsums4})] and solving for $k$, provides a
discontinuous cross-over porosity $f_r(k)$ curve which defines in
the $(f,k)$ plane boundary lines between dilute and high-anisotropy
regions. Owing to the approximations at play, this boundary cannot
be trusted for $k$ of order one (for this reason we do not display
the curves). On the other hand, we find $f_r(k)\simeq
(9\pi^3/1024)k\simeq k/\pi$ for $k\ll 1$ and $f_r(k)\simeq
(9\pi^3/512)k^{-1}\simeq 2/(k \pi)$ for $k\gg 1$.

Due to the relation $f=\pi a^2$, the cross-over porosity in the
highly anisotropic regime stems from a length scale $\xi(k)$ such
that $\xi\sim a/k^{1/2}$ for $k\ll 1$, and $\xi\sim a k^{1/2}$ for
$k\gg 1$. From a mathematical standpoint, these length scales
originate from a scaling property of the lattice sums. We focus here
on the case $k\to 0$. The case $k\to\infty$ can be discussed by
adapting this argument. Introducing $K=k/(1-k)$, the sum $S_\mu$ in
(\ref{eq:ssums}) can be written with a summand proportional to $[1+K
\gamma(\mathbf{\hat p})]^{-1}$, where the dimensionless quantity
$\gamma(\mathbf{\hat p})$ reads
\begin{equation}
\gamma(\mathbf{\hat p})=\frac{(p_x^2+p_y^2)^2}{4 p_x^2 p_y^2}.
\end{equation}
Singling out the contribution of the main diagonal to $S_\mu$, the
remainder of this sum can be brought down to a sum over $p_x\geq 2$
and $1\leq p_y\leq p_x-1$, in which $1/4 \leq \gamma(\mathbf{\hat
p}) \leq (p_x/1)^2$. Hence, $\gamma(\mathbf{\hat p})\sim p^2$ so
that $K \gamma(\mathbf{\hat p})$ provides an appreciable
$k$-dependent contribution only for $p\gtrsim 1/\sqrt{K}\sim
1/\sqrt{k}$. Moreover, $[2 J_1(x)/x]^2$ is appreciable only when
$x\lesssim 2$. In terms of $p$, this reads $p\lesssim 1/(\pi a)$,
see (\ref{eq:ssums}). Hence $k$-dependent terms contribute only
provided that $1/\sqrt{k}\leq p \leq 1/(\pi a)$. In turn, this is
possible only if $\xi(k)\lesssim 1$. For $\xi(k)\gtrsim 1$, a
$k$-independent regime instead takes place in $S_\mu$.

From a physical standpoint, the length scale $\xi$ represents an
\emph{effective inclusion size}. Fig.\ \ref{fig:chgmtregFFT} indeed
displays three maps of the parallel strain field in PS loading,
computed by FFT at fixed anisotropy ratio $k=10^{-3}$ with varying
porosity $f\simeq f_r/10$, $f_r$ and $10 f_r$. It is seen that
localized shear bands develop from the void as porosity increases.
At regime change, they coalesce and span the entire medium. The void
can be considered as an isolated inclusion only for $f<f_r$. A
similar effect takes place for high $k$ values. We checked
numerically that in both cases, before coalescence, the strain
intensity in the bands decays exponentially as $\varepsilon\propto
\exp(- b\, r/\xi)$, where $r$ is the distance from the void, and
where $b$ is a numerical coefficient of order one.

\subsection{\label{sec:afsd}Average fields and standard deviations}
\subsubsection{General considerations}
The first two moments of the fields are required for applications to
non-linear EMAs, and can be consistently computed from any linear
homogenization estimate.\cite{PONT98} Hereafter,
$\overline{\varepsilon}_e^{(\alpha)}\equiv \left\langle
\varepsilon\right\rangle^{(\alpha)}/\overline{\varepsilon}$ denotes
the phase average of a strain component $\varepsilon$, normalized by
the applied macroscopic field. Likewise, we denote by
$SD^{(\alpha)}(\varepsilon)$ its standard deviation (SD) in phase
$\alpha$ , \emph{normalized by $\overline{\varepsilon}$}. Similar
notations apply to stress components.

The phase-averaged fields in the porous composite are deduced from
the set of equations
\begin{subequations}
\begin{eqnarray}
\label{eq:poroussigav}
\widetilde{\mathbb{L}}:\overline{\bvarepsilon}&=&(1-f)\mathbb{L}^{(1)}:\langle\bvarepsilon\rangle^{(1)},\\
\overline{\bvarepsilon}&=&(1-f)\langle\bvarepsilon\rangle^{(1)}+f\langle\bvarepsilon\rangle^{(2)}.
\end{eqnarray}
\end{subequations} Moreover, assuming single mode-loading, the second
moments in each phase are obtained by taking a derivative of the
strain energy with respect to the elastic moduli of the phases,
as\cite{PONT98}
\begin{eqnarray}
\label{eq:fluctder}
\left\langle\varepsilon_{\text{m},\text{SS},\text{PS}}^2\right\rangle^{(\alpha)}
=\frac{1}{c^{(\alpha)}} \frac{\partial \widetilde{L}}{\partial
L^{(\alpha)}}
\langle\varepsilon_{\text{m},\text{SS},\text{PS}}\rangle^2,
\end{eqnarray}
where $\widetilde{L}$ is $\widetilde{\kappa}$ (resp.\
$\widetilde{\lambda}$, $\widetilde{\mu}$) when the index in the
l.h.s.\ is $m$ (resp.\ $SS$, $PS$) and where $L^{(\alpha)}$ is
$\kappa^{(\alpha)}$ (resp.\ $\lambda^{(\alpha)}$, $\mu^{(\alpha)}$)
when the index the r.h.s.\ is $m$ (resp.\ $SS$, $PS$). The variances
follow. If need be, the incompressibility limit is taken after these
quantities are computed.

Table \ref{fig:MomentsVSkSimpleShear} displays for $f=0.1$ the
normalized phase-average strains
$\overline{\varepsilon}_e^{(\alpha)}$ for $\alpha=1$, $2$ and SDs of
the strain and stress components in the matrix, as computed by the
EMA and by full-field calculations. The overall agreement is again
excellent, the most important observed deviations, if any, occurring
at small $k$. The table layout emphasizes the qualitative
correspondence between case $(k,SS)$ and case $(1/k,PS)$, explained
in Ref.\ \onlinecite{WILL07}.

Some trends in the data are explained by appealing to the
variational expression of the elastic energy $W$:
\begin{equation}
W(\overline{\varepsilon};k;f) = \inf_{\varepsilon\in
\mathcal{K}(\overline{\varepsilon})}
      \left\lbrace \frac{1}{2}\int \varepsilon:\mathbb{L}:\varepsilon \right\rbrace\\
 = \frac{1}{2}\overline{\varepsilon}:\widetilde{\mathbb{L}}:\overline{\varepsilon}
\label{eq:energy}
\end{equation}
where $\mathcal{K}(\overline{\varepsilon})=\{\varepsilon;
\varepsilon_{ij}=(\partial_i u_j+\partial_j u_i)/2,
 \left\langle\varepsilon\right\rangle=\overline{\varepsilon}\}$
is the set of admissible strain fields. E.g., for an incompressible
material under SS loading (\ref{eq:poroussigav}) and
(\ref{eq:energy}) imply:
\begin{equation}\label{eq:flucta0}
\overline{\varepsilon}\,\langle \varepsilon_\parallel\rangle^{(1)}
=\langle \varepsilon_\parallel^2\rangle^{(1)}+(1/k)
 \langle \varepsilon_\perp^2\rangle^{(1)}.
\end{equation}
Hence the standard deviation $SD^{(1)}(\varepsilon_\parallel)$ of
the parallel component of the strain is essentially finite, since
$\smash{\langle \varepsilon_\parallel\rangle^{(1)}}$ is, in
agreement with the analytical expressions of the SDs in the next
section to which we refer the reader for this discussion. Consider
now another strain field $\varepsilon'$, solution for an anisotropy
ratio $k'>k$. Using it as a trial field for problem
(\ref{eq:energy}) with $k$ provides one inequality. Duplicating the
argument with $k$ and $k'$, and $\varepsilon$, $\varepsilon'$
interchanged, yields after some easy algebra involving
(\ref{eq:flucta0}):
\begin{equation}\label{eq:ineq}
SD^{(1)}(\varepsilon_\perp)^2\leq \frac{\langle
\varepsilon'_\parallel\rangle^{(1)}-\langle
\varepsilon_\parallel\rangle^{(1)}}{(1/k')-(1/k)}\leq
SD^{(1)}(\varepsilon'_\perp)^2,
\end{equation}
which entails (\ref{eq:fluctder}) for $k'\to k$. Thus, the standard
deviation $SD^{(1)}(\varepsilon_\perp)$ of the transverse (PS)
component of the strain field increases with $k$ at $f$ fixed,
consistently with Table \ref{fig:MomentsVSkSimpleShear}. Moreover,
using (\ref{eq:ineq}) and the equality $\partial_k
\langle\varepsilon_\perp^2\rangle^{(1)}=-k
\partial_k \langle\varepsilon_\parallel^2\rangle^{(1)}$ [from (\ref{eq:fluctder})]
shows that under SS loading
$\smash{\langle\varepsilon_\parallel^2\rangle^{(1)}}$ is a
decreasing function of $k$. These considerations hold for any fixed
microstructure.

Analyzing FFT calculations at $f=0.1$ for various values of $k$ in
log-log plots (not shown), we observe that (for this $f$) the SDs
behave as powers of $k$ with numerical exponents close to $1/4$ or
$3/4$: e.g., under SS loading, $SD^{(1)}(\varepsilon_\perp)$ decays
as $k^{3/4}$ when $k\to 0$, and blows up as $k^{1/4}$ when
$k\to\infty$; meanwhile,
$SD^{(1)}(\sigma_\perp)=SD^{(1)}(\varepsilon_\perp)/k\sim k^{-1/4}$
as $k\to 0$ and $\sim k^{-3/4}$ as $k\to\infty$. The ``soft'' case
$k\to 0$ is in agreement with the dilute analytical expressions
(\ref{eq:sssdeps}) and (\ref{eq:sssdsps}) below, which indicates
that the computed systems remained in the dilute regime $f\ll
f_r(k)\sim k$. On the other hand, the ``hard'' case $k\to \infty$
where strong strain localization takes place (see map C in Table
\ref{tab:displacement}) is consistent with (\ref{eq:sssdeps}) and
(\ref{eq:sssdsps}) only if we replace $f$ by $f_r(k)\sim k^{-1}$ in
these expressions. Thus, here, $SD^{(1)}(\varepsilon_\perp)$ blows
up [see (\ref{eq:sssdeps2})], but behaves as though the system
remained in the cross-over regime. This information, extracted
numerically, is not contained in the expressions
(\ref{eq:sssdeps3}), (\ref{eq:pssdess3}), for which we could only
produce limiting values.

Actually, in the limit $k\to\infty$, infinite SDs in the
\emph{transverse} component of the strain result from its
concentration as Dirac lines (see Sec.\ \ref{sec:dispmaps}), and are
linked to discontinuities (jumps) in its \emph{parallel}
component.\cite{WILL07} This results in a deformation pattern by a
``rigid block sliding'' mechanism, the ``rigid blocks'' being here
connected parts of matter separated by discontinuity lines. This
block-sliding effect only takes place provided that the strain jump
lines have ``percolated''. Below ``percolation'', sliding is
impossible in a linear material and the transverse strain
fluctuations described by (\ref{eq:sssdeps}) strongly increase with
$k$ as $SD^{(1)}(\varepsilon_\perp)\sim k^{3/4}$. On the contrary,
beyond ``percolation'', sliding takes place and
$SD^{(1)}(\varepsilon_\perp)\sim k^{1/4}$ increases in a weaker way,
since sliding makes deformation easier. Analogous properties are
found under PS loading, provided that $k$ is replaced by $1/k$:
e.g., $SD^{(1)}(\varepsilon_\perp)$ is a decreasing function of $k$
and blows up in the hard loading mode as $\sim k^{-1/4}$ when $k\to
0$.

Table \ref{fig:MeanStrainVSc} shows numerical results for the strain
and stress field averages and SDs, plotted for various anisotropy
ratios $k=0.01$, $1$ and $100$. EMA estimates are provided for
comparisons. Except when SDs blow up at strong anisotropy, the EMA
estimates are in good agreements with FFT results, for porosities up
to $f\sim 0.4$. It is worth observing that, in situations of high
anisotropy ratios, a change in the structure of the strip patterns
in the material coincides with a change of concavity of the standard
deviations $SD^{(1)}(\varepsilon_\parallel)$ of the parallel
component of the strain field. For instance, when SS loading is
applied at $k\gg 1$, the bands cover the whole medium at $f\approx
\pi/8\approx 0.4$. Around this value, the quantity
$SD^{(1)}(\varepsilon_\parallel)$ changes from a concave to a convex
function of $f$. Such a change also occurs at $f\approx \pi/8$ for
the same SDs when $k\gg 1$ and PS loading is applied. FFT field maps
then indicate that the structure of the strain pattern also
undergoes an abrupt change at this point (with the appearance of
thinner strips linking closest neighboring voids -- not shown).

\begin{table*}
\caption{\label{fig:MomentsVSkSimpleShear} \emph{SS and PS
loadings}. Comparisons between EMA estimates (solid lines) and FFT
results (dots) at porosity $f=0.1$, for averages of the strain along
the loading direction in each phase, and standard deviations (SD) of
stress and strain components in the matrix, vs.\ matrix anisotropy
ratio $k=\lambda/\mu$. Strains and stresses are normalized by the
appropriate macroscopic component in the loading direction
(macroscopic strain $\overline{\epsilon}=\langle\epsilon\rangle$, or
stress $\overline{\sigma}=\langle\sigma\rangle$). SDs in the voids
are irrelevant.}
\includegraphics[width=17.94cm]{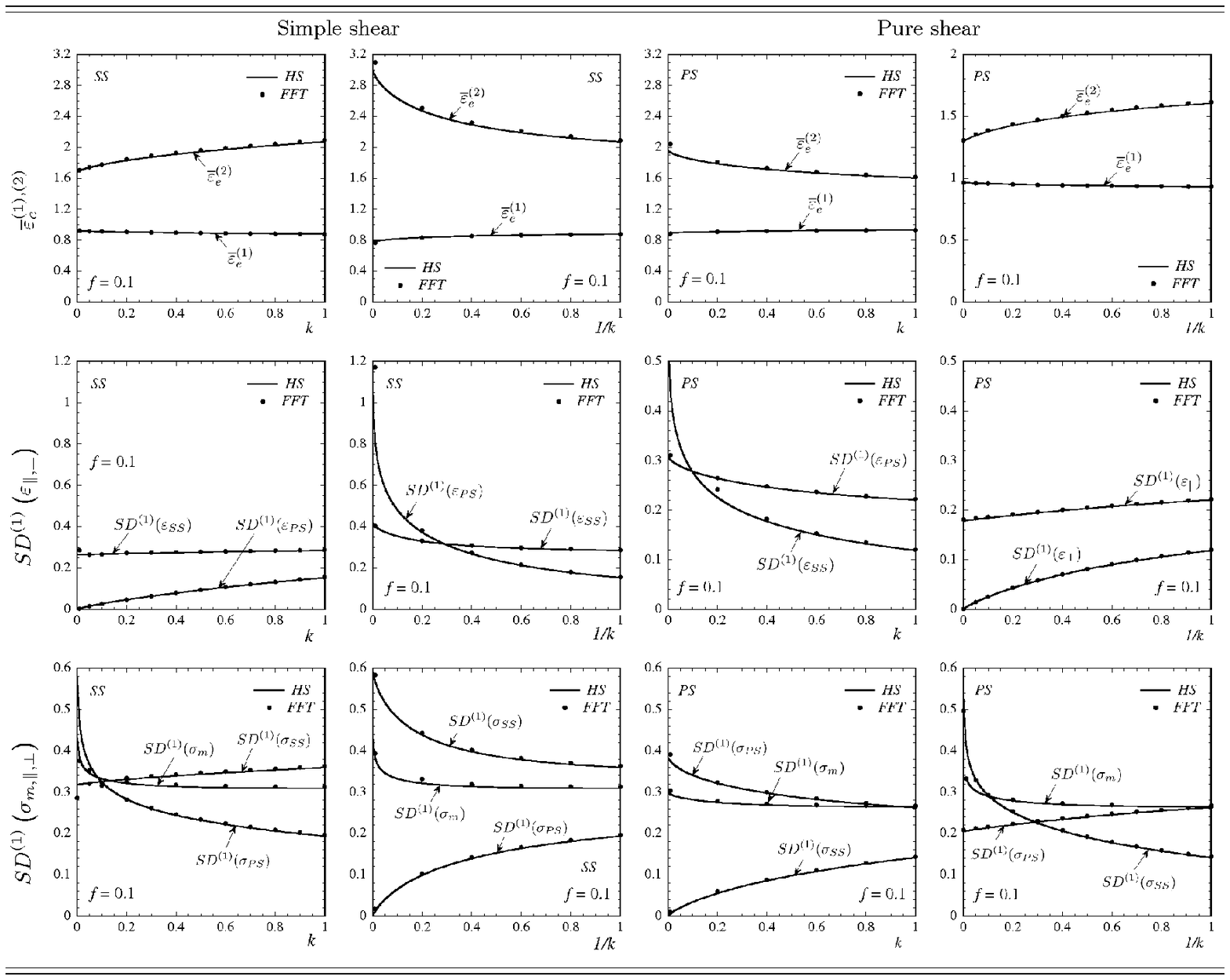}
\end{table*}
%
\begin{turnpage}
\begin{table*}
\caption{\label{fig:MeanStrainVSc} Average shear strains
$\overline{\varepsilon}_e^{(\alpha)}$ in the matrix ($\alpha=1$) and
in voids ($\alpha=2$) vs.\ porosity $f$. Comparisons between EMA
estimates (solid lines), FFT results (dots) and exact analytical
results at $k=0$, $\infty$(dash-dotted lines) for particular values
of the matrix anisotropy ratio $k=\lambda/\mu$, in pure shear (PS)
and simple shear (SS) loadings. The normalization is the same as for
Table III (see legend).}
\includegraphics*[height=23.50cm,angle=-90]{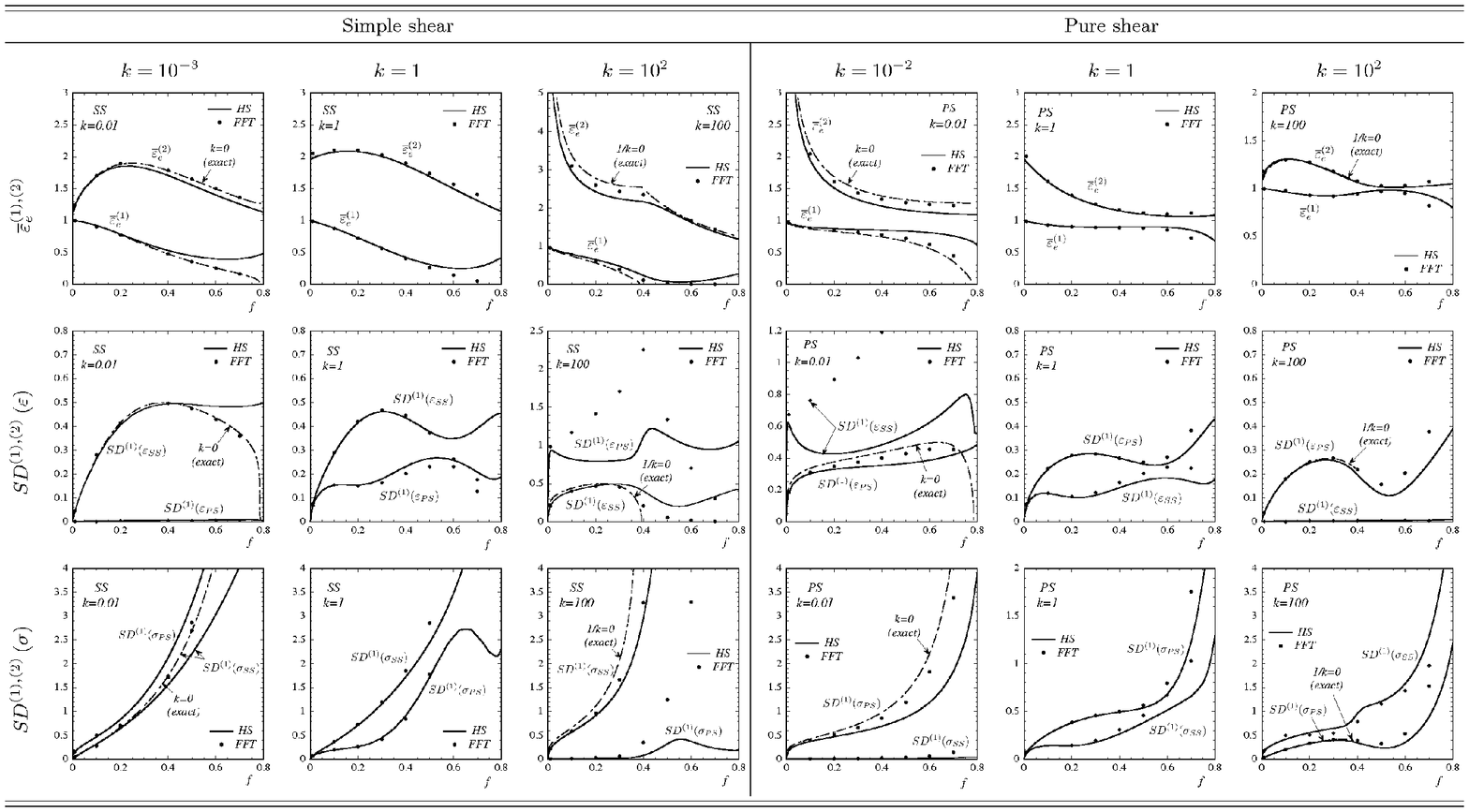}
\end{table*}
\end{turnpage}
%
\subsubsection{Standard deviations in the dilute limit}
\label{sec:sddilute} For completeness, we include the standard
deviations in the dilute limit (of relevance to nonlinear EMAs)
computed from the HS-EMA.

For SS loading, we find for $f\ll f_r(k)$ (low anisotropy or low
porosity),
\begin{subequations}
\begin{eqnarray}
SD(\varepsilon_\parallel)&\sim& SD(\sigma_\parallel)\sim
 f^{1/2}k^{1/4}, \label{eq:sssdess}\\
SD(\varepsilon_\perp)&\sim& f^{1/2}k^{3/4},\label{eq:sssdeps}
\\
 SD(\sigma_\perp)&\sim & f^{1/2}k^{-1/4},\label{eq:sssdsps}
\end{eqnarray}
\end{subequations}
where the $f^{1/2}$ proportionality of the SDs goes along
``classical'' $O(f)$ dilute corrections to the effective moduli. In
limits of infinite anisotropy, where necessarily $f\gg f_r(k)=0$,
the HS estimates provide:
\begin{subequations}
\begin{eqnarray}
\label{eq:sssdess2} && \hspace{-2em}SD(\varepsilon_\parallel) =
SD(\sigma_\parallel) = \frac{4\sqrt{2}}{\pi^{3/4}\sqrt{3}}f^{3/4}
\nonumber\\&& +\left(\frac{64\sqrt{2}}{3\pi^{9/4}\sqrt{3}}
-\frac{\sqrt{3}\pi^{3/4}}{4\sqrt{2}}\right)f^{5/4},
\quad k\to 0,\\
&&SD(\varepsilon_\perp) = 0, \quad SD(\sigma_\perp) = \infty, \quad k\to 0,\\
&&SD(\varepsilon_\parallel) = SD(\sigma_\parallel) = \frac{\pi^{3/4}\sqrt{3}}{2^{9/4}}f^{1/4},
\quad\!\! k\to\infty,\\
\label{eq:sssdeps2} &&SD(\varepsilon_\perp) = \infty, \quad
SD(\sigma_\perp) = 0, \quad k\to \infty,
\end{eqnarray}
\end{subequations}
while exact results read:\cite{WILL07}
\begin{subequations}
\begin{eqnarray}
&&SD(\varepsilon_\parallel)= SD(\sigma_\parallel) =
  \frac{4\sqrt{2}}{\sqrt{3}\pi^{3/4}}f^{3/4}
\nonumber\\&&
 +\frac{\sqrt{3}\pi^{3/4}}{8\sqrt{2}}\left(\frac{6}{\pi}+\frac{8}{\pi^2}-1\right)f^{5/4},
\quad k\to 0, \label{eq:num1} \\
&&SD(\varepsilon_\perp)=0, \quad SD(\sigma_\perp) = \infty, \quad k\to 0, \\
&&SD(\varepsilon_\parallel)= SD(\sigma_\parallel) =\left(2f/\pi\right)^{1/4},\quad k\to \infty, \\
\label{eq:sssdeps3}
&&SD(\varepsilon_\perp)=\infty,\quad
SD(\sigma_\perp) = 0,\quad k\to \infty.
\end{eqnarray}
\end{subequations}

For PS loading, the HS-EMA gives for $f\ll f_r(k)$:
\begin{subequations}
\begin{eqnarray}
SD(\varepsilon_\parallel)&\sim& SD(\sigma_\parallel) \sim f^{1/2}k^{-1/4}, \label{eq:pssdeps} \\
SD(\varepsilon_\perp)&\sim& f^{1/2}k^{-3/4},\label{eq:pssdess} \\
SD(\sigma_\perp) &\sim& f^{1/2}k^{1/4}. \label{eq:pssdsps}
\end{eqnarray}
\end{subequations}
For infinite anisotropy where $f\gg f_r(k)=0$, the HS estimates
provide:
\begin{subequations}
\begin{eqnarray}
\label{eq:pssdeps2}&&SD(\varepsilon_\parallel) =
SD(\sigma_\parallel) = \frac{\pi^{3/4}\sqrt{3}}{4\sqrt{2}}f^{1/4},
\quad\!\! k\to 0,\\
&&SD(\varepsilon_\perp) = \infty, \quad SD(\sigma_\perp) = 0, \quad k\to 0,\\
&&
\!\!\!\!\!SD(\varepsilon_\parallel) = SD(\sigma_\parallel) =
\frac{2^{9/4}}{\sqrt{3}\pi^{3/4}}f^{3/4}
\nonumber\\&&
+\left(\frac{2^{3/4}32}{3^{3/2}\pi^{9/4}}-\frac{\sqrt{3}\pi^{3/4}}{2^{9/4}}\right)f^{5/4},
\quad k\to \infty,\\
&&SD(\varepsilon_\perp) = 0, \quad SD(\sigma_\perp) = \infty, \quad
k\to \infty,
\end{eqnarray}\end{subequations}
while exact results are:\cite{WILL07}
\begin{subequations}
\begin{eqnarray}
&&SD(\varepsilon_\parallel)= SD(\sigma_\parallel) =\left(f/\pi\right)^{1/4},\quad k\to 0, \\
\label{eq:pssdess3}
&&SD(\varepsilon_\perp)=\infty, \quad SD(\sigma_\perp) = 0, \quad k\to 0, \\
&&SD(\varepsilon_\parallel)= SD(\sigma_\parallel) =
  \frac{2^{9/4}}{\sqrt{3}\pi^{3/4}}f^{3/4}
\nonumber\\&&
 +\frac{\sqrt{3}\pi^{3/4}}{2^{1/4}8}\left(\frac{3}{\pi}+\frac{4}{\pi^2}-1\right)f^{5/4},
\quad k\to \infty, \\
&&SD(\varepsilon_\perp)=0,\quad SD(\sigma_\perp) = \infty,\quad k\to \infty.
\end{eqnarray}
\end{subequations}
Thus, the HS-EMA correctly reproduces the scaling behavior of the
SDs in all cases.

\section{Concluding discussion}
To summarize, we compared the results of the HS-EMA to FFT
calculations, and showed that the agreement is excellent, even in
the non-trivial case of localizing behavior, as far as effective
moduli and averaged fields are concerned, and provided that the void
concentration lies below $0.3$. This result is relevant to the study
of \emph{non-linear} effective-medium techniques: the latter
involving both an anisotropic EMA, and a specific self-consistent
linearization procedure (which determines the effective anisotropy
of the former), the present study shows that should strong
deviations between FFT and EMA results in nonlinear media be
observed (in the similar set-up of a periodic voided medium, and in
similar conditions of porosity and of effective anisotropy), they
ought be attributed to the linearization procedure rather than to
the underlying liner EMA, even in limits of high effective
anisotropy (determined by the field fluctuations in the nonlinear
theory). Also, the present work provides a useful independent
confirmation of the involved analytical analysis of Ref.\
\onlinecite{WILL07}.

As a by-product of the study, of possible practical applications, we
showed that by combining a regular lattice of voids (which makes the
structure lighter) and an anisotropic matrix, properties could be
tuned so as to make the overall medium elastically isotropic in
plane strain.

We also studied analytically the lattice sums which underly the EMA
approach, and showed that they possess a scaling property which, in
the dilute limit of small porosity and at high (but finite)
anisotropy, allows for a cross-over between regular and singular
porosity dependence of the effective medium. A length scale $\xi$
was associated to this scaling, and interpreted as an effective
heterogeneity size. It mathematically diverges in the limit of
infinite anisotropy. However, its physically associated counterpart
being constrained by the finite size of the cell in the periodic
medium, cross-over occurs when the effective heterogeneities
``percolate'', i.e.\ when $\xi$ is trivially of order one. This
corresponds to a strongly correlated regime of strain localization
bands spanning the system.

Actually, scaling properties of lattice sums similar to the one
considered here, have already been pointed out by
Barber,\cite{BARB77} elaborating on Hall's work,\cite{HALL73} in a
purely mathematical context (in particular, no explanation in terms
of length scales was given). Here, we make a connection between this
phenomenon and strain localization in anisotropic elastic media.
Barber's paper also provides a means to compute the cross-over
function. However, our lattice sums lead to technical difficulties
which preclude the straightforward obtention of a similar result,
and we leave this issue for future work.

Moreover, we found that under loading in a ``hard" direction of the
anisotropic medium, standard deviations of the transverse strain
component blow up continuously as a power of the anisotropy ratio,
as though the system remained in a cross-over regime. This absence
of finite threshold for diverging fluctuations, and the
above-described behavior, suggest the existence of a special type of
continuous phase transition, of infinite order (called a ``weak
phase transition'' by Hall,\cite{HALL73}), here obviously without
symmetry breaking. The presence of logarithmic terms in $k$
(identified numerically in the Appendix) also hints in this
direction, since logarithmic corrections to scaling constitute a
hallmark of transitions of infinite order.\cite{WEIG05} However, a
random version of the system should be investigated before reaching
definite conclusions.

Finally, it was observed in Ref.\ \onlinecite{WILL07} that the
singular effective moduli in the limit of infinite anisotropy are
directly connected to the hyperbolic character of the governing
equations in this limit. The very existence of a cross-over shows
without ambiguity that the problem, although elliptic from a strict
mathematical point of view, presents a ``quasi-hyperbolic''
character at short distances for high but finite anisotropy. This
observation may be of relevance to theoretical investigations of
granular materials, for which a model with a similar anisotropic
matrix has been proposed.\cite{OTTO03}

\begin{acknowledgments}
The work of M.I.I.\ and P.P.C.\ was supported by NSF grant
CMS-02-01454. The work of F.W.\ was supported by a CEA Ph.D.\ grant.
We gratefully thank P.\ Suquet for having kindly provided to us the
notes of Ref.\ \onlinecite{SUQU91}.
\end{acknowledgments}

\appendix*
\section{\label{sec:latticesums}Asymptotics of lattice sums and dilute limits}
We extract the dilute expansions of the lattice sums
$S_{\lambda,\mu}$ when $k\to 0,\infty$ as follows. With the notation
\begin{equation}
A(a,x)=[2 J_1(2\pi a x)/x\,]^2,
\end{equation}
write the lattice sums, with $p=(p_x^2+p_y^2)^{1/2}$, as
\begin{eqnarray}
S_\lambda&=&\frac{1}{\pi}\hspace{-1em}\sum_{\atop{p_x\geq 0}{p_y\geq
1, p_y\not=p_x}}
A(a,p)\, \left(p_x^2-p_y^2\right)^2 /D(\mathbf{p}),\\
S_\mu&=&\frac{1}{\pi}\sum_{\atop{p_x\geq 1}{p_y\geq 1}} A(a,p)\, 4
p_x^2 p_y^2 /D(\mathbf{p}),
\end{eqnarray}
where $D(\mathbf{p})=4 p_x^2 p_y^2+k(p_x^2-p_y^2)^2$. The above
expressions explicitly acknowledge the fact that the principal
diagonal $p_x=p_y$ does not contribute to $S_\lambda$, and that the
Cartesian axis $p_x=0$ does no contribute to $S_\mu$. Introducing
the lattice sums
\begin{equation}
S_1(a)=\frac{1}{\pi}\sum_{n\geq 1}A(a,n),\quad
S_2(a)=\frac{1}{\pi}\sum_{\atop{p_x\geq 0}{p_y\geq 1}}A(a,p),
\end{equation}
provides for $k\to \infty,0$,
\begin{eqnarray}
k
S_\lambda^{k\to\infty}&=&\frac{1}{\pi}\hspace{-1em}\sum_{\atop{p_x\geq
0}{p_y\geq 1,
p_y\not=p_x}}A(a,p)=S_2(a)-\frac{1}{2}S_1(\sqrt{2}a),\nonumber\\
\label{eq:smul} S_\mu^{k\to 0}&=&\frac{1}{\pi}\sum_{\atop{p_x\geq
1}{p_y\geq 1}}A(a,p)=S_2(a)-S_1(a).
\end{eqnarray}
In  $S_\lambda^{k\to\infty}$ (resp.\ $S_\mu^{k\to 0}$) the
contribution of the principal diagonal (resp.\ Cartesian axis
$p_x=0$) has been subtracted from $S_2(a)$. The factors $\sqrt{2}$
and $1/2$ result from $n$ being replaced by $\sqrt{2}n$ on the main
diagonal.

One privileged tool for exact asymptotic expansions is the Mellin
transform.\cite{WONG89} The transform and its inverse are defined
by:
\begin{eqnarray*}
M[f(x);z]&=&\int_0^\infty {\text{d}}x\, x^{z-1}f(x),\\
f(x)&=&\frac{1}{2i\pi}\int_{c-i\infty}^{c+i\infty} {\text{d}}z\,
x^{-z} M[f(x);z],
\end{eqnarray*}
where $c$ lies within the analyticity strip (parallel to the
imaginary axis) of $M[f(x);z]$ in the $z$-plane. Shifting the
inversion contour to the left encircles the poles on the negative
$z$ axis and provides the asymptotic series expansion around $x=0$
in positive powers of $x$. Conversely, shifting the contour to the
right provides the asymptotic expansion near $x=\infty$ in negative
powers of $x$. The Mellin transform provides, for $0<c<3$,
$$
A(a,p)=\frac{1}{2i\pi}\hspace{-0.3em}\int_{c-i\infty}^{c+i\infty}\hspace{-0.7em}
\frac{{\rm d}z}{(2\pi
a)^z}\frac{1}{p^z}\frac{2}{\sqrt{\pi}}\frac{\Gamma(z/2)\Gamma(3/2-z/2)}{\Gamma(2-z/2)\Gamma(3-z/2)}
$$
Next appealing to the definition of the Zeta function for $z>1$,
$\zeta(z)=\sum_{n\geq 1} 1/n^z$, and to Hardy's lattice
sum\cite{HARD19,GLAS73}
\begin{equation}
\label{eq:hardy} \sum_{\atop{p_x\geq 0}{p_y\geq
1}}\frac{1}{(p_x^2+p_y^2)^{z/2}}=\zeta(z/2)\beta(z/2),\quad (z>2)
\end{equation}
where $\beta(z)=\sum_{n\geq 0} (-1)^n/(2n+1)$ is the Dirichlet (or
Catalan) function,\cite{GLAS73} and interchanging the lattice sums
and the contour integral yields:
\begin{subequations}
\begin{eqnarray}
S_1(a)&=&\frac{-i}{\pi^{5/2}}\hspace{-0.3em}\int_{c_1-i\infty}^{c_1+i\infty}\hspace{-0.7em}
\frac{{\rm d}z}{(2\pi
a)^z}\zeta(z)\frac{\Gamma(z/2)\Gamma(3/2-z/2)}{\Gamma(2-z/2)\Gamma(3-z/2)},\nonumber\\
\\
S_2(a)&=&\frac{-i}{\pi^{5/2}}\hspace{-0.3em}\int_{c_2-i\infty}^{c_2+i\infty}\hspace{-0.7em}
\frac{{\rm d}z}{(2\pi
a)^z}\zeta(z/2)\beta(z/2)\nonumber\\
&&\hspace{2cm}\times
\frac{\Gamma(z/2)\Gamma(3/2-z/2)}{\Gamma(2-z/2)\Gamma(3-z/2)},
\end{eqnarray}
\end{subequations}
where $1<c_1<3$ and where $2<c_2<3$ as the result of the above
restrictions. The following properties hold: $\Gamma(z)$ has simple
poles at negative integers $z=-k\leq 0$ and has no zeros; $\zeta(z)$
has only one simple pole at $z=1$, and has (so-called ``trivial'')
zeros at even, nonzero, negative integers; $\beta(z)$ has no poles
and has zeros at odd negative integers. Then, by shifting the
contour to the left in both integrals, only the poles $z=1$, $0$,
contribute to $S_1$, and only the poles at $z=2$ and $z=0$
contribute to $S_2$. Eventually we obtain:
\begin{subequations}
\label{eq:asympt}
\begin{eqnarray}
\label{eq:asympt1}
S_1(a)&=&\frac{32}{3\pi}a-2\pi a^2,\\
\label{eq:asympt2} S_2(a)&=&1-\pi a^2.
\end{eqnarray}
\end{subequations}
The polynomial form of these expressions indicates that they are
actually \emph{exact}, since for such functions the asymptotic
expansion coincides with the function itself.
\begin{figure}[ht]
\includegraphics[height=7cm,angle=-90]{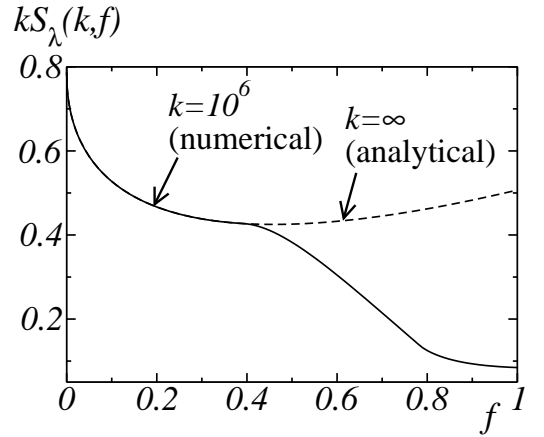}
\caption{\label{fig:sumSl} Quantity $kS_\lambda(k,a)$ vs.\ void
concentration $f=\pi a^2$ as $k\to\infty$: comparison between
(\ref{eq:limitsums3}), dashed line, and a numerical calculation of
the lattice sum, solid line, for $k=10^6$. Breakdown of
(\ref{eq:limitsums3}) occurs at $f=\pi/8\simeq 0.39$.}
\end{figure}

The validity of (\ref{eq:asympt1}) is linked to the decay of
$$
f(z)=\left|s^{-z}\zeta(z)\frac{\Gamma(z/2)\Gamma(3/2-z/2)}{\Gamma(2-z/2)\Gamma(3-z/2)}\right|
\quad\text{as } \Re{\rm e}z\to-\infty.
$$
We have in the limit $f(z)$ $\sim
s^{-z}|\zeta(z)\Gamma(z/2)\zeta(z)$ $/\Gamma(-z/2)|$. Owing to the
reflection formula
$$
\pi^{-z/2}\Gamma(z/2)\zeta(z)=\pi^{-(1-z)/2}\Gamma\bigl((1-z)/2\bigr)\zeta(1-z),
$$
$|\Gamma(z/2)\zeta(z)/\Gamma(-z/2)|\sim \pi^z |\zeta(-z)| \sim
\pi^z$. Hence the contribution of the integration line in the limit
$c_1\to-\infty$ is negligible only if $s<\pi$. In terms of $a$, this
amounts to $a<1/2$. The breakdown of the obtained expressions thus
corresponds to the close-packing limit $a=1/2$. A similar reasoning
using the corresponding reflection formula for $\beta(z)$ provides
the same range of validity for $S_2$.

Combining  (\ref{eq:smul}), (\ref{eq:asympt}) and
(\ref{eq:relation}) then results in
\begin{subequations}
\label{eq:limitsums}
\begin{eqnarray}
\label{eq:limitsums1}
k S_\lambda^{k\to 0}&=&\frac{32}{3\pi}a-2\pi a^2,\quad (a<1/2)\\
\label{eq:limitsums2}
S_\mu^{k\to 0}&=&1-\frac{32}{3\pi}a+\pi a^2,\quad (a<1/2)\\
\label{eq:limitsums3}
k S_\lambda^{k\to\infty}&=&1-\frac{16\sqrt{2}}{3\pi}a+\pi a^2,\,\, (a<{\scriptstyle\frac{1}{2\sqrt 2}})\\
\label{eq:limitsums4}
S_\mu^{k\to\infty}&=&\frac{16\sqrt{2}}{3\pi}a-2\pi a^2,\,\,
(a<{\scriptstyle\frac{1}{2\sqrt{2}}}).
\end{eqnarray}
\end{subequations}
The restrictions $a<1/2$ and $a<1/(2\sqrt{2})$ correspond to
critical concentrations $f=\pi/4$ and $f=\pi/8$. At these points,
either the voids percolate ($k=0$, PS or SS and $k=\infty$, SS) or
the shear bands undertake a configurational change ($k=\infty$,
PS).\cite{WILL07} An illustration of the breakdown of expression
(\ref{eq:limitsums3}) is provided by Fig.\ \ref{fig:sumSl}.

We could not compute analytically the leading corrections in $k$ to
these sums. However, by carefully analyzing brute force numerical
computations of the sums for $k$ down to $10^{-5}$ or up to $10^6$
for $f=0.1$, we found that corrections to (\ref{eq:limitsums1}),
(\ref{eq:limitsums2}), (\ref{eq:limitsums3}), (\ref{eq:limitsums4})
are of the form $O(k\log k)$, $O(-k\log k)$, $O\bigl(k^{-1}\log
k\bigr)$, $O\bigl(-k^{-1}\log k\bigr)$, respectively.

As a final remark, we emphasize that only (divergent) asymptotic
series for $S_1(a)$, $S_2(a)$ at $a>1/2$ can be obtained: then, the
integrand in both contour integrals blows up as $[z/(4 \pi a e)]^z$
where $e$ is Euler's constant. These asymptotic expansions are
easily extracted. We do not provide them here since the region
$a\gtrsim 1/2$ cannot be examined without appealing to additional
investigation procedures (e.g., Pad\'e approximants) which lie
outside the scope of this paper.

\begin{thebibliography}{99}
\bibitem{NOTE2} Specifically, $E=\mu(3-\mu/\kappa)$, $\nu=(\kappa-\mu)/(2\kappa)$ for plane
strain; $E=4\kappa\mu/(\kappa+\mu)$, $\nu=(\kappa-\mu)/(\kappa+\mu)$
for plane stress; the shear modulus $\lambda$ of the simple shear
mode is unchanged.
\end{thebibliography}
\end{document}